\documentclass{emulateapj}

\tabletypesize{\scriptsize}
\usepackage{url}
\usepackage{multirow}
\usepackage{epstopdf}

\newcommand{\DataRunningDays}{0.008}
\newcommand{\DataRunningMinutes}{11.5}
\newcommand{\INSERTNIGHTSSIMPJBAND}{15}	%Total number of nights of J-band observations
\newcommand{\INSERTSPREADDAYSSIMPJBAND}{90}		%Total days from first to last observation of the J-band observations
\newcommand{\INSERTNIGHTSSIMP}{17}	%Total number of nights of observations
\newcommand{\INSERTSPREADDAYSSIMP}{122}		%Total days from first to last observation
\newcommand{\TOTALNONOVERLAPPINGHOURS}{124}			%Total Number of non-overlapping hours in our photometry

\newcommand{\INSERTNIGHTSTVLM}{3}	%Total number of nights of observations
\newcommand{\INSERTNIGHTSSIMULTANEOUSTVLM}{2}	%Total number of simultaneous nights of observations
\newcommand{\INSERTSPREADDAYSTVLM}{46}		%Total days from first to last observation
\newcommand{\PeriodHoursSIMPZeroOneThreeSixJband}{2.48}
\newcommand{\PeriodHoursErrorSIMPZeroOneThreeSixJband}{0.10}

\newcommand{\PeriodHoursSIMPZeroOneThreeSixKsbandFifteenDecemberSeventeen}{2.55}
\newcommand{\PeriodHoursErrorSIMPZeroOneThreeSixKsbandFifteenDecemberSeventeen}{0.12}

\newcommand{\PeriodHoursFifteenDecemberSeventeenKsBand}{2.55}
\newcommand{\PeriodHoursErrorFifteenDecemberSeventeenKsBand}{0.07}
\newcommand{\WeightedMeanSIMPPeriod}{2.406}
\newcommand{\WeightedErrorSIMPPeriod}{0.004}
\newcommand{\WeightedErrorSIMPPeriodScaled}{0.008}
\newcommand{\SIMPPeriodScaleFactor}{2}
\newcommand{\DiffRotationReducedChiSquaredComment}{would be close to one}
\newcommand{\ReducedChiSquaredDiffRotation}{1.8}
\newcommand{\DiffRotationPercentage}{2}
\newcommand{\SIMPPeriodScaleFactorFifteenNovemberThirtyFactor}{2}
\newcommand{\SIMPMaximumRot}{6}	%In excess of
\newcommand{\SIMPMinimumRot}{1}	%Less than...
\newcommand{\PeriodHoursTVLMJband}{1.99}
\newcommand{\PeriodHoursErrorTVLMJband}{0.04}
\newcommand{\SinusoidPeaktoPeakAmplitudePercentageTVLMSixteenMarchThreeOnezband}{0.37}
\newcommand{\SinusoidPeaktoPeakAmplitudePercentageErrorTVLMSixteenMarchThreeOnezband}{0.20}

\newcommand{\SinusoidPhaseTVLMSixteenMarchThreeOnezband}{0.056}
\newcommand{\SinusoidPhaseErrorTVLMSixteenMarchThreeOnezband}{1.056}

\newcommand{\SinusoidPeaktoPeakAmplitudePercentageTVLMSixteenMayFifteenJband}{0.72}
\newcommand{\SinusoidPeaktoPeakAmplitudePercentageErrorTVLMSixteenMayFifteenJband}{0.11}

\newcommand{\SinusoidPhaseTVLMSixteenMayFifteenJband}{0.088}
\newcommand{\SinusoidPhaseErrorTVLMSixteenMayFifteenJband}{0.026}

\newcommand{\SinusoidPeaktoPeakAmplitudePercentageTVLMSixteenMayFifteenIband}{0.27}
\newcommand{\SinusoidPeaktoPeakAmplitudePercentageErrorTVLMSixteenMayFifteenIband}{0.20}
\newcommand{\SinusoidPeaktoPeakAmplitudePercentageThreeSigmaTVLMSixteenMayFifteenIband}{0.90}
\newcommand{\SinusoidPhaseTVLMSixteenMayFifteenIband}{0.262}
\newcommand{\SinusoidPhaseErrorTVLMSixteenMayFifteenIband}{1.068}

\newcommand{\SinusoidPeaktoPeakAmplitudePercentageTVLMSixteenMaySixteenIband}{0.16}
\newcommand{\SinusoidPeaktoPeakAmplitudePercentageErrorTVLMSixteenMaySixteenIband}{0.11}
\newcommand{\SinusoidPeaktoPeakAmplitudePercentageThreeSigmaTVLMSixteenMaySixteenIband}{0.67}
\newcommand{\SinusoidPhaseTVLMSixteenMaySixteenIband}{0.352}
\newcommand{\SinusoidPhaseErrorTVLMSixteenMaySixteenIband}{0.818}

\newcommand{\SinusoidPeaktoPeakAmplitudePercentageTVLMSixteenMaySixteenJband}{0.95}
\newcommand{\SinusoidPeaktoPeakAmplitudePercentageErrorTVLMSixteenMaySixteenJband}{0.06}

\newcommand{\SinusoidPhaseTVLMSixteenMaySixteenJband}{0.058}
\newcommand{\SinusoidPhaseErrorTVLMSixteenMaySixteenJband}{0.009}

\shorttitle{Multiwavelength Photometry of Ultra-Cool Dwarfs} 
\shortauthors{Croll et al.}

\begin{document}

\title{Long-term, Multiwavelength Light Curves of Ultra-Cool Dwarfs: II. The evolving Light Curves of the T2.5 SIMP 0136 \& the Uncorrelated Light Curves of the
M9 TVLM 513}

\author{
	Bryce Croll\altaffilmark{1},
	Philip S. Muirhead\altaffilmark{1} \altaffilmark{2},	
	Jack Lichtman\altaffilmark{3}, 
 	Eunkyu Han\altaffilmark{2},
	Paul A. Dalba\altaffilmark{2},
%	Brian Taylor\altaffilmark{22}
	Jacqueline Radigan\altaffilmark{4}
}

\altaffiltext{1}{Institute for Astrophysical Research, Boston University, 725 Commonwealth Ave. Room 506, Boston, MA 02215; croll@bu.edu}

\altaffiltext{2}{Department of Astronomy, Boston University, 725 Commonwealth Ave., Boston, MA 02215, USA}

% \altaffiltext{22}{Insert Address}

\altaffiltext{3}{Department of Physics, University of Connecticut, 2152 Hillside Road, Unit 3046 Storrs, CT 06269-3046}

\altaffiltext{4}{Utah Valley University, Orem, UT 84058, USA}

\begin{abstract}
 
We present  \INSERTNIGHTSSIMP \ nights of ground-based,
near-infrared photometry of the variable L/T transition brown dwarf SIMP J013656.5+093347 and an additional
\INSERTNIGHTSTVLM \ nights of ground-based photometry of the radio-active late M-dwarf TVLM 513-46546.
Our TVLM 513-46546 photometry includes 
\INSERTNIGHTSSIMULTANEOUSTVLM \ nights of simultaneous, multiwavelength, ground-based photometry, in which we detect
obvious J-band variability, but do not detect I-band variability of similar amplitude, confirming
that the variability
of TVLM 513-46546 most likely arises from clouds or aurorae, rather than starspots.
Our photometry of SIMP J013656.5+093347 includes \INSERTNIGHTSSIMPJBAND \ nights of J-band photometry
that allow us to observe how the variable light curve of this L/T transition brown dwarf
evolves from rotation period to rotation period, night-to-night and week-to-week.
We estimate the rotation period of SIMP J013656.5+093347 as 
\WeightedMeanSIMPPeriod \ $\pm$ \WeightedErrorSIMPPeriodScaled \ hours, and do not find evidence for obvious differential rotation.
The 
peak-to-peak amplitude 
displayed by SIMP J013656.5+093347 in our light curves evolves from greater than \SIMPMaximumRot\% to less than \SIMPMinimumRot\% in a matter of days,
and the typical timescale for significant evolution of the SIMP J013656.5+093347 light curve 
appears to be approximately $<$1 to 10 rotation periods.
This suggests that those 
performing spectrophotometric observations
of brown dwarfs should be cautious in their interpretations comparing the spectra between a variable brown dwarf's maximum flux and minimum flux
from observations lasting only approximately a rotation period,
as these comparisons may depict the spectral 
characteristics 
of a single, ephemeral snapshot, rather than the full range of characteristics.

\end{abstract}


\section{Introduction}

Over the last number of years simultaneous, multiwavelength photometric and spectrophotometric observations of 
variable brown dwarfs at 
the L/T transition have enabled a number of detailed investigations attempting to elucidate the cause of the observed variability
(e.g. \citealt{Artigau09,Radigan12,Buenzli12,Apai13,Buenzli14,Buenzli15,Yang15,Yang16}).
The result of these investigations has been increased understanding of the role that fluctuations in cloud condensate opacity play
in the observed
variability. 
Upon the first detections of significant variability at the L/T transition (e.g. \citealt{Artigau09,Radigan12}) the belief 
was that the observed fluctuations were caused by gaps or holes in thick clouds
exposing hotter regions underneath, rotating in and out of view \citep{AckermanMarley01,Burgasser02,Artigau09}.
More recent simultaneous multiwavelength photometry, or spectrophotometry -- 
especially driven by the impressive, spectrophotometric capabilities of the WFC3 instrument \citep{MacKenty10} on the Hubble Space
Telescope ({\it HST}) --
have revealed the possibility that these variations were more specifically caused by thick or thin clouds \citep{Radigan12,Buenzli15}, rather than outright cloud-free regions, 
or caused by varying haze layers \citep{Yang15}.

 However, a number of these detailed studies only observed these variable brown dwarfs for tens of minutes
(e.g. \citealt{Buenzli14}), to hours at a time (e.g. \citealt{Apai13,Buenzli15,Yang15}).
This is in spite of the fact that, since the first detections of large amplitude variability in L/T transition brown dwarfs,
the photometric light curves of some of 
these objects have been shown to display significant evolution in their light curves (e.g. \citealt{Artigau09,Radigan12,Gillon13}).
Thus, the conclusions these detailed, spectrophotometric studies reach are based inherently on an ephemeral snapshot, that might not be 
indicative of the usual behaviour of the L/T dwarf, or might not cover the extreme range of variability behaviour displayed by the ultra-cool dwarf.

To determine whether these detailed, spectrophotometric observations capture the range of characteristics of ultra-cool dwarfs,
longer-term light curves of these variable ultra-cool dwarfs are required. To date, a handful of nights of photometry have captured the evolution of the light curves of the 
L/T transition brown dwarf
SIMP J013656.5+093347 (e.g. \citealt{Artigau09,Metchev13,Radigan14}),
and the significant night-to-night evolving variability of 
the L/T transition Luhman-16 binary system \citep{Gillon13}.
{\it Kepler} has now returned photometry of two early L dwarfs allowing the evolution, or lack thereof, of their optical
light curves to be tracked for
two years for an L0 dwarf (WISEP J190648.47+401106.8; \citealt{Gizis13,Gizis15}) and $\sim$36 days for an L8 dwarf (WISEP J060738.65+242953.4; \citealt{Gizis16}).
Relatively constant variability was detected for the L0 dwarf WISEP J190648.47+401106.8 \citep{Gizis13,Gizis15},
while for the L8 dwarf WISEP J060738.65+242953.4 a lack of variability was detected, arguably
due to a pole-on inclination of the ultra-cool dwarf \citep{Gizis16}.
What is needed to advance this science is many more long term light curves of ultra-cool dwarfs that accurately track the evolution, or lack thereof, of
the light curves of these objects from rotation period to rotation period, night-to-night, week-to-week, season-to-season and even year-to-year.

 In addition to L/T transition objects, the cause of variability of M/L transition objects has also attracted growing
interest recently (e.g. \citealt{Gelino02,Littlefair08,Harding13,Metchev15,Hallinan15}). 
Starspots are the usual explanation for variability of ultra-cool dwarfs on the stellar side of 
the hydrogen-fusing limit. M-dwarfs are notoriously active -- nearly all
very late M-dwarfs are active, with detections of H$\alpha$,
a common activity marker, rising throughout the M-spectral class (\citealt{West04}; \citealt{Schmidt15}).
Rotation periods for M-dwarfs, revealed by Doppler imaging techniques
(\citealt{BarnesCollier01} ; \citealt{Barnes04}),
and by photometry (\citealt{Rockenfeller06}; \citealt{Irwin11}),
have indicated that starspots are
pervasive on M-dwarfs.

Recently, clouds have been found to play a role not only on L/T transition objects, but possibly throughout the whole L spectral class \citep{Metchev15}.
Therefore, it is possible that clouds may even lead to some of the observed variability for very late M-dwarfs.  
The detection of anti-correlated light curves on the M9 dwarf TVLM 513-46546 was 
at first inferred to be due to the presence of clouds even on this late M-dwarf \citep{Littlefair08}.

More recently, in addition to clouds, aurorae -- 
similar to the planetary aurorae in our solar system (e.g. \citealt{Clarke80}) -- have been suggested to be another possibility to explain the 
observed variability of ultra-cool dwarfs.
Such auroral activity has been observed on an M8.5 dwarf at the end of the main sequence \citep{Hallinan15}. Multiwavelength photometry of this dwarf displays
light curves that are anti-correlated in phase, and \citet{Hallinan15} speculate that the aforementioned anti-correlated light curves of TVLM 513-46546 \citep{Littlefair08},
and possibly other late M-dwarfs,
may result from auroral activity. In addition, auroral activity may not simply be constrained to the M/L transition, as recent
radio detections of polarized, pulsed emissions from a T2.5 dwarf \citep{Kao16} and a T6.5 dwarf \citep{RouteWolszczan12,WilliamsBerger15} indicate
this phenomenon may extend even to the L/T transition.
Therefore, an additional possibility is that the variability at the L/T transition is caused in part or wholly by auroral activity,
rather than the commonly accepted
fluctuations in cloud condensate opacity.

%%%%%%%%%%%%%%%%%%%%%%%%%%%%%%%%%%%%%%%%%%%%%%%%%%%%%%%%
%not sure the best placement of this sentence if warranted at all
%Observing the ultra-cool dwarf at two or more bands simultaneously can be done in order to determine if the variability is caused
%by sunspots (\citet{Rockenfeller06}) or by dust clouds (\citet{Littlefair06}). 
%%%%%%%%%%%%%%%%%%%%%%%%%%%%%%%%%%%%%%%%%%%%%%%%%%%%%%%

Furthermore, it is also possible, or arguably likely, that the variability of a single ultra-cool dwarf may be driven by more than a single one 
of the previously mentioned astrophysical variability mechanisms. 
Magnetically driven cool or hot starspots may be periodically obscured by time evolving clouds (\citealt{Lane07}; \citealt{Heinze13}; \citealt{Metchev15}),
or the variability of predominantly cloudy
brown dwarfs may be affected by the presence of occasional aurorae  \citep{Hallinan15}. Longer term light curves of ultra-cool dwarfs can also inform our understanding
of whether one or more of these physical mechanisms play a role in the observed variability on a single ultra-cool dwarf. 

% It may be due to a single variability changing method that these brown dwarfs have this rapid evolution - such as 
% decreasing cloud thickness, or the starspots sizes growing- or due to a secondary variability mechanism in addition to the primary mechanism. Therefore it is imperative to monitor
% ultra-cool dwarfs for multiple rotation cycles spread over days, weeks, months, and possibly even years.
% - Talk about aurorae (seen in M-dwarfs, but does it play a role in L/T transition variability).

% - Provide a summary of the various sections

To illuminate these various questions here we attempt to observe how the variability of two well known variable ultra-cool dwarfs
evolves from rotation period to rotation period, night-to-night and week-to-week.
In Section \ref{SecUCDTargets} we provide an overview of our two targets: the variable T2.5 brown dwarf SIMP J013656.5+093347,
and the variable M9 dwarf TVLM 513-46546.
In Section \ref{SecObs} we present
\INSERTNIGHTSSIMP \ nights of photometry of SIMP J013656.5+093347,
and \INSERTNIGHTSTVLM \ nights of photometry of TVLM 513-46546, including 
\INSERTNIGHTSSIMULTANEOUSTVLM \ nights of simultaneous multiwavelength photometry of TVLM 513-46546.
In Section \ref{SecAnalysisSIMP} we analyze the rapidly evolving light curves of SIMP J013656.5+093347;
for SIMP J013656.5+093347 the peak-to-peak amplitude of variability evolves from a minimum of less than \SIMPMinimumRot\% to
a maximum of greater than \SIMPMaximumRot\% in just a handful of nights.
This suggests that those performing detailed comparisons of the spectra of L/T transition 
brown dwarfs between the maximum flux and minimum flux displayed in a 
observation lasting only a single rotation period,
should be aware that this comparison may reveal very different conclusions if the variability amplitude
is small (less than $\sim$\SIMPMinimumRot\% peak-to-peak amplitudes),
compared to when the variability is considerably greater.
In Section \ref{SecAnalysisTVLM} we show that the fact we detect near-infrared variability without detectable accompanying optical
variability from our multiwavelength photometry confirms that the variability of TVLM 513-46546 most likely arises
from clouds or aurorae.

\section{Our Ultra-Cool Dwarf Targets}
\label{SecUCDTargets}

\subsection{The T2.5 dwarf SIMP 0136}

The T2.5 dwarf SIMP J013656.5+093347 (hereafter SIMP 0136; \citealt{Artigau06})
is one of the best studied variable brown dwarfs at the L/T transition.
It is arguably the archetype for L/T transition brown dwarfs displaying significant variability,
as it was the first brown dwarf discovered that exhibited large amplitude variability; 
\citet{Artigau09} first announced that 
SIMP 0136 displayed prominent $\sim$5\% variability from a number of nights of J-band monitoring
of this dwarf
using the 1.6-m Observatoire du Mont Megantic.
The \citet{Artigau09} SIMP 0136 light curves displayed obvious evolution from night-to-night, 
as some nights of J-band monitoring displayed quasi-sinusoidal variability,
while others displayed double-peaked behavior indicative of at least two spot/cloud groups separated in longitude
on the brown dwarf.
In spite of the obvious evolution from night-to-night of the SIMP 0136 light curves,
\citet{Artigau09} estimated the rotation period of SIMP 0136 to be 
2.3895 $\pm$ 0.0005 hours from fitting two sinusoids to their
two nights of double-peaked, ground-based light curves.

Despite the obvious, large amplitude variability displayed by SIMP 0136, the number of nights of
photometric follow-up to observe how the light curve of SIMP 0136 evolves from rotation period
to rotation period, night-to-night, and season to season has been limited. 
\citet{Metchev13} presented an additional 6 nights of J-band monitoring of SIMP 0136 from 
the 1.6-m Observatoire du Mont Megantic.
\citet{Apai13} obtained six {\it HST} orbits of SIMP 0136 utilizing HST/WFC;
they found the spectra were best-fit by a combination of thick, dimmer, low-temperature
clouds and thin, brighter, warm clouds.
\citet{Radigan14} observed SIMP 0136 in J-band for $\sim$2.3 hours and detected
2.9\% peak-to-peak variability.
\citet{Yang16} performed simultaneous {\it HST}/J-band and Spitzer/IRAC 3.6 $\mu m$, and some
4.5 $\mu m$, photometry of SIMP 0136 for a few hours
and detected a $\sim$30 degree phase shift
between their J-band and 3.6 $\mu m$ photometry;
\citet{Yang16} confirmed that SIMP 0136 variability was driven by clouds,
with different layers of clouds
at different pressure levels, and therefore depths in the atmosphere, causing the observed phase shifts.
Most recently, \citet{Kao16} detected SIMP 0136 at radio wavelengths, although they were not able to determine a radio period.

Here we attempt to probe the evolution of the light curve of SIMP 0136 
by presenting 
\INSERTNIGHTSSIMPJBAND \ nights of J-band photometric monitoring of SIMP 0136
spread out over \INSERTSPREADDAYSSIMPJBAND\ days using the 1.8-m Perkins Telescope at Lowell Observatory.
Our many nights of J-band observations allow us to observe how the variability of SIMP 0136
evolves from rotation period to rotation period, night-to-night and week-to-week.
Specifically, we demonstrate that the peak-to-peak amplitude of variability
evolves from less than \SIMPMinimumRot\% to
greater than \SIMPMaximumRot\% for a rotation period in approximately a week.

\subsection{The M9 dwarf TVLM 513-46546}

%Williams et al. 2015 has a good summary that we can copy/adapt from
The rapidly-rotating, radio active M9 dwarf TVLM 513-46546 (hereafter TVLM 513) has 
undergone considerable monitoring at radio wavelengths and occasional monitoring at 
optical wavelengths.
Although TVLM 513 was originally assigned a spectral type of M8.5 \citep{Kirkpatrick95}, more recently
\citet{Reid08} and \citet{West11} assign a spectral type of M9 to TVLM 513, and
\citet{Dahn02} report an effective temperature of $\sim$2300 K.
TVLM 513 is most likely a star at the very bottom of the main sequence \citep{Tinney93};
a lack of lithium absorption from TVLM 513 suggests a mass in excess of $\sim$0.06 $M_{\odot}$, and an age
greater than $\sim$100 Myr \citep{Martin94,Reid02}.

%Many more references in Williams et al. 2015 for radio wavelengths...
TVLM 513 has been a frequent target for monitoring at radio wavelengths (e.g. \citealt{Berger02,Hallinan07,Berger08}).
\citet{Hallinan07} detected prominent radio bursts from TVLM 513 with a period of $\sim$ 1.96 hours
that they inferred to be the rotation period of this ultra-cool dwarf.
Emission from TVLM 513 has also been detected at millimeter wavelengths \citep{Williams15Mm} with 
the Atacama Large Millimeter/submillimeter Array (ALMA), 
a detection that is inferred to be due to the synchrotron process.
%The H-alpha emission has been described as relatively weak, although I'd need to know about H-alpha widths...
H$\alpha$ emission has also been detected from TVLM 513 \citep{Martin94,Basri01}.

TVLM 513 has also undergone significant optical photometric monitoring of its variability.
\cite{Lane07} detected I-band variability at the $\sim$10 mmag level from 12 hours of optical monitoring,
and suggested this variability was driven by magnetic starspots on this ultra-cool dwarf.
\citet{Harding13} presented 10 nights of I and i'-band monitoring of TVLM 513 with
peak-to-peak variability amplitudes that varied between $\sim$0.6 - 1.2\% from their photometry;
\citet{Harding13} used
these light curves to infer a rotation period of 1.95958 $\pm$ 0.00005 hr.
Perhaps most intriguingly, \citet{Littlefair08} performed
2.7 hours of simultaneous optical monitoring of TVLM 513 in the
%Sloan \textit{g}' and Sloan \textit{i}' bands
g' and i'-bands
and detected anti-correlated 3-4\% sinusoidal variability
with a period of approximately 2 hours.
\citet{Littlefair08} initially interpreted this anti-correlated light curve to likely be the result of 
inhomogeneous cloud coverage on TVLM 513; 
however, more recently, with the detection of auroral emission on another similar
late M-dwarf (the M8.5 dwarf LSR J1835+3259; \citealt{Hallinan15}), 
auroral emission is now another plausible explanation for the
observed anti-correlated variability.
More recently, \citet{MilesPaez15} performed optical linear polarimetry of TVLM 513 and detected $\sim$2 hour 
periodicity in their data.

%INSERTHEREe H-alpha

%In observing simultaneously in the 
%two bands \citet{Littlefair08} found that the variability in the Sloan \textit{g}' and Sloan \textit{i}' bands to
%be anti-correlated.  Due to this anti-correlation among other explanaitions \citet{Littlefair08} considered star spots
%to be an unlikely explanation for the variability in the dwarf. 

 Here we present 
\INSERTNIGHTSTVLM \ nights of photometry of TVLM 513, including
\INSERTNIGHTSSIMULTANEOUSTVLM \ nights 
of simultaneous optical and near-infrared photometric monitoring of this ultra-cool dwarf. Our
simultaneous, multiwavelength photometry of TVLM 513 displays prominent J-band variability with a $\sim$2.0 hour
period, similar to the previous radio detections of this object. As we are unable to detect prominent
I-band variability from our simultaneous optical photometry for this object, the most likely explanation
for the variability of TVLM 513 is either clouds or aurora, rather than starspots.

\section{Observations}
\label{SecObs}

\begin{deluxetable*}{ccccccccc}
\tablecaption{SIMP 0136 Observing Log}
\tablehead{
\colhead{Date} 		& \colhead{Telescope}		& \colhead{Observing}	& \colhead{Duration\tablenotemark{a}}	& \colhead{Exposure} 	& \colhead{Overhead\tablenotemark{b}}	& \colhead{Airmass}				& \colhead{Conditions}	& \colhead{Aperture \tablenotemark{c}}\\
\colhead{(UTC)}		& \colhead{\& Instrument}	& \colhead{Band}	& \colhead{(hours)}			& \colhead{Time (sec)}	& \colhead{(sec)}			& \colhead{}					& \colhead{}		& \colhead{(pixels)}\\
}
\centering
\startdata
2015/08/18		& Perkins/MIMIR			& J			& 3.23 					& 30			& 3	& 1.34 $\rightarrow$ 1.11 $\rightarrow$ 1.13	& Clear			& 5, 15, 25  \\	%
2015/08/21		& Perkins/MIMIR			& J			& 3.21 					& 30			& 3	& 1.29 $\rightarrow$ 1.11 $\rightarrow$ 1.14	& Clear			& 6, 15, 25  \\	%
2015/09/27		& Perkins/PRISM			& z'			& 7.75 					& 150			& 18	& 1.83 $\rightarrow$ 1.11 $\rightarrow$ 2.01	& Clear			& 6, 15, 27  \\	%
2015/10/09		& Perkins/MIMIR			& J			& 2.87 					& 30			& 8	& 1.17 $\rightarrow$ 2.14  			& Clear			& 6, 15, 25  \\	%
2015/11/10		& Perkins/MIMIR			& J			& 8.82 					& 30			& 3	& 2.61 $\rightarrow$ 1.11 $\rightarrow$ 2.18	& Clear and windy 	& 6, 15, 25  \\	%
2015/11/12		& Perkins/MIMIR			& J			& 9.47 					& 30			& 3	& 2.62 $\rightarrow$ 1.11 $\rightarrow$ 3.00	& Clear			& 6, 15, 25  \\	%
2015/11/13		& Perkins/MIMIR			& J			& 8.82 					& 30			& 3	& 2.34 $\rightarrow$ 1.11 $\rightarrow$ 2.40	& Thin cirrus		& 6, 15, 25  \\	%
2015/11/18		& Perkins/MIMIR			& J			& 8.71 					& 30			& 3	& 2.12 $\rightarrow$ 1.11 $\rightarrow$ 2.56	& Occasional clouds	& 6, 15, 25  \\	%
2015/11/19		& Perkins/MIMIR			& J			& 7.96 					& 30			& 3	& 1.99 $\rightarrow$ 1.11 $\rightarrow$ 2.00	& Occasional clouds	& 6, 15, 25  \\	%
2015/11/20		& Perkins/MIMIR			& J,Ks			& 8.56 					& 30,15			& 3,3	& 2.01 $\rightarrow$ 1.11 $\rightarrow$ 2.57	& Clear			& 6 [7]\tablenotemark{d}, 15, 25  \\	%
2015/11/21		& Perkins/MIMIR			& J			& 8.74 					& 30			& 3	& 2.01 $\rightarrow$ 1.11 $\rightarrow$ 2.78	& Occasional clouds	& 6, 15, 25  \\	%
2015/11/22		& Perkins/MIMIR			& J			& 8.61 					& 30			& 3	& 2.05 $\rightarrow$ 1.11 $\rightarrow$ 2.54	& Clear			& 6, 15, 25  \\	%
2015/11/25		& Perkins/MIMIR			& J			& 5.41 					& 30			& 3	& 1.76 $\rightarrow$ 1.11 $\rightarrow$ 1.23	& Occasional clouds	& 6, 15, 25  \\	%
2015/11/27		& Perkins/MIMIR			& J			& 8.78 					& 30			& 3	& 1.86 $\rightarrow$ 1.11 $\rightarrow$ 3.26	& Clear			& 6, 15, 25  \\	%
%%%% 2015/11/28		& Perkins/MIMIR			& J			&  					& 30			& 3	&  $\rightarrow$  $\rightarrow$ 		& ?			& ?, ?, ?  \\	%	Too little data to analyze this evening, ignore...
2015/11/29		& Perkins/MIMIR			& J			& 8.10 					& 30,25			& 3	& 1.78 $\rightarrow$ 1.11 $\rightarrow$ 2.42	& Clear			& 6, 15, 25  \\	%
%I don't really have sheets for this..., probably in my gmail, actually a 9.2 second overhead
2015/11/29		& Hall/NASA42			& z'			& 7.92 					& 180			& 9	& 1.76 $\rightarrow$ 1.11 $\rightarrow$ 2.27	& Clear			& 4, 14, 26  \\	%
2015/11/30		& Perkins/MIMIR			& J			& 8.46					& 30			& 3	& 1.78 $\rightarrow$ 1.11 $\rightarrow$ 2.90	& Clear			& 6, 15, 25  \\	% Brian Skiff observed this evening, and it is not clear to me the observing log survived. In my version of the log sheet it doesn't say that it was cloudy, so let's assume it was clear.
2015/12/17		& Perkins/MIMIR			& Ks			& 6.42					& 15			& 3	& 1.26 $\rightarrow$ 1.11 $\rightarrow$ 2.38	& Clear			& 6, 15, 25  \\	% 

\enddata
\tablenotetext{a}{The duration indicates the time between the first and last observation of the evening, and does not take into account gaps in the data due to clouds, humidity or poor weather.}
\tablenotetext{b}{The overhead includes time for read-out, and any other applicable overheads.}
\tablenotetext{c}{We give the radius of the aperture, the radius of the inner annulus and the radius of the outer annulus that we use for sky subtraction in pixels.}
\tablenotetext{d}{6 is the radius of the aperture for our J-band analysis, while 7 is the aperture radius for our Ks-band analysis from this night.}
\label{TableSIMPObs}
\end{deluxetable*}

\begin{deluxetable*}{ccccccccc}
\tablecaption{TVLM 513 Observing Log}
\tablehead{
\colhead{Date} 		& \colhead{Telescope}		& \colhead{Observing}	& \colhead{Duration\tablenotemark{a}}	& \colhead{Exposure} 	& \colhead{Overhead\tablenotemark{b}}	& \colhead{Airmass}				& \colhead{Conditions}	& \colhead{Aperture \tablenotemark{c}}\\
\colhead{(UTC)}		& \colhead{\& Instrument}	& \colhead{Band}	& \colhead{(hours)}			& \colhead{Time (sec)}	& \colhead{(sec)}			& \colhead{}					& \colhead{}		& \colhead{(pixels)}\\
}
\centering
\startdata
2016/05/15		& Perkins/MIMIR			& J			& 7.28 					& 10			& 3	& 1.27 $\rightarrow$ 1.02 $\rightarrow$ 1.96	& Initial light clouds	& 6, 15, 25 	\\	%
2016/05/15		& Hall/42			& I			& 8.16 					& 180			& 9	& 1.54 $\rightarrow$ 1.02 $\rightarrow$ 1.93	& Initial light clouds	& 7, 14, 26 	\\	%
2016/05/16		& Perkins/MIMIR			& J			& 8.69					& 10			& 3	& 1.59 $\rightarrow$ 1.02 $\rightarrow$ 2.27	& Clear			& 6, 15, 25 	\\	%
2016/05/16		& Hall/NASA42			& I			& 8.04 					& 180			& 9	& 1.49 $\rightarrow$ 1.02 $\rightarrow$ 1.95	& Clear			& 4.5, 14, 26	\\	%
\enddata
\tablenotetext{a}{The duration indicates the time between the first and last observation of the evening, and does not take into account gaps in the data due to clouds, humidity or poor weather.}
\tablenotetext{b}{The overhead includes time for read-out, and any other applicable overheads.}
\tablenotetext{c}{We give the radius of the aperture, the radius of the inner annulus and the radius of the outer annulus that we use for sky subtraction in pixels.}
\label{TableTVLMObs}
\end{deluxetable*}

\begin{figure*}
% \centering
\includegraphics[scale=0.50, angle = 270]{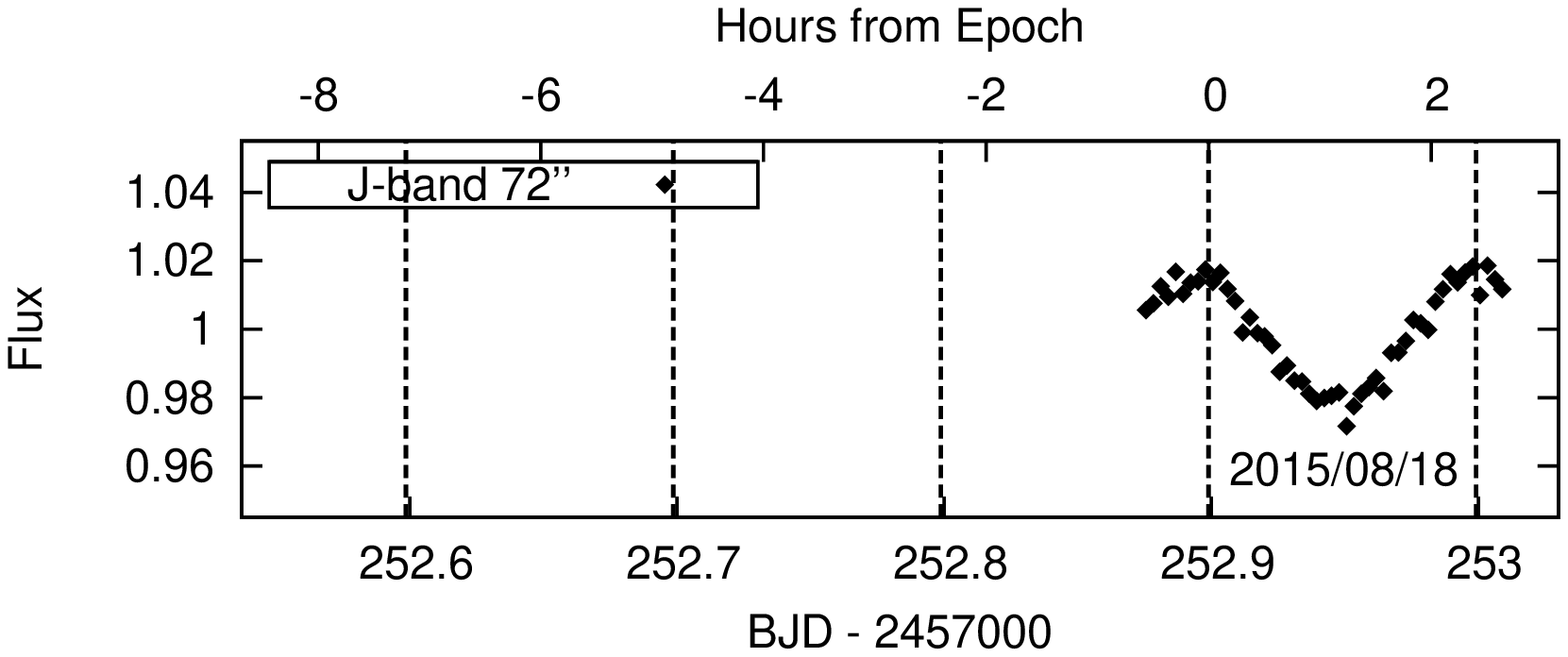}
\includegraphics[scale=0.50, angle = 270]{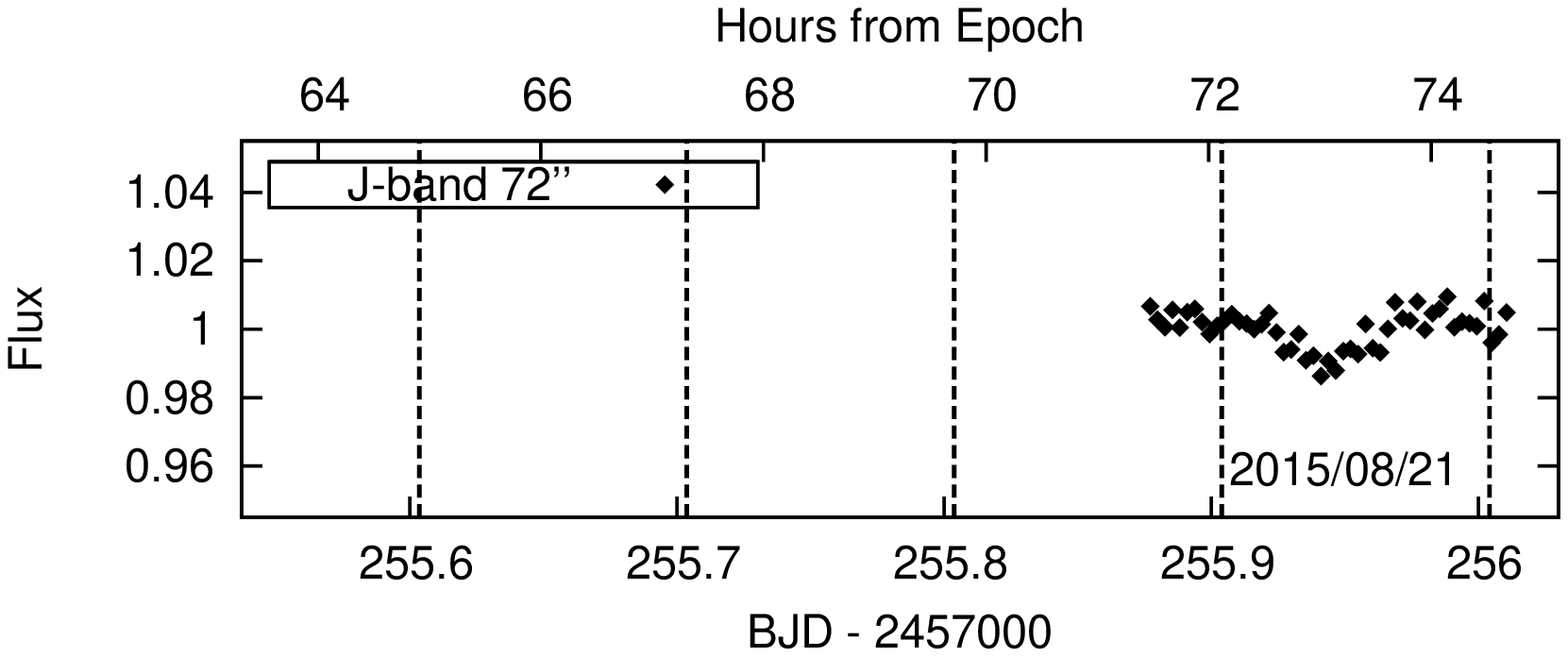}
\includegraphics[scale=0.50, angle = 270]{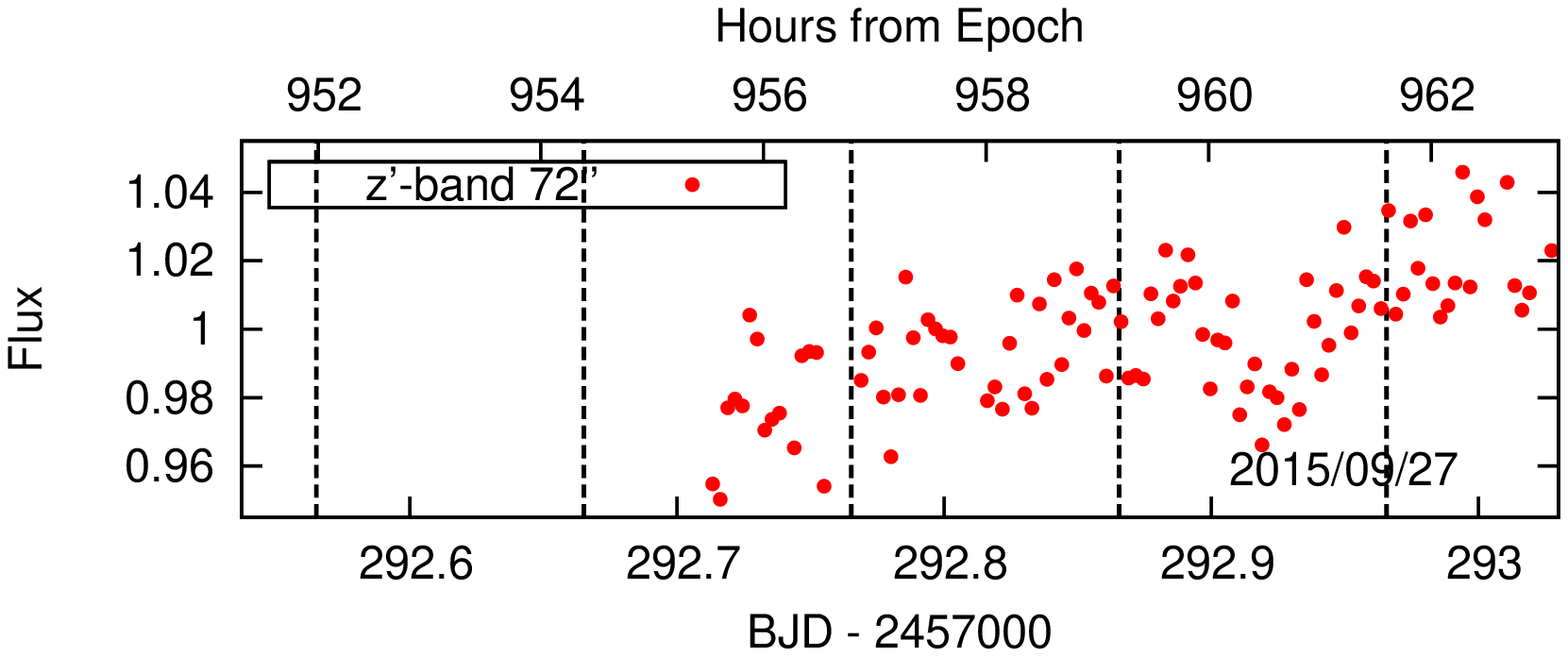}
\includegraphics[scale=0.50, angle = 270]{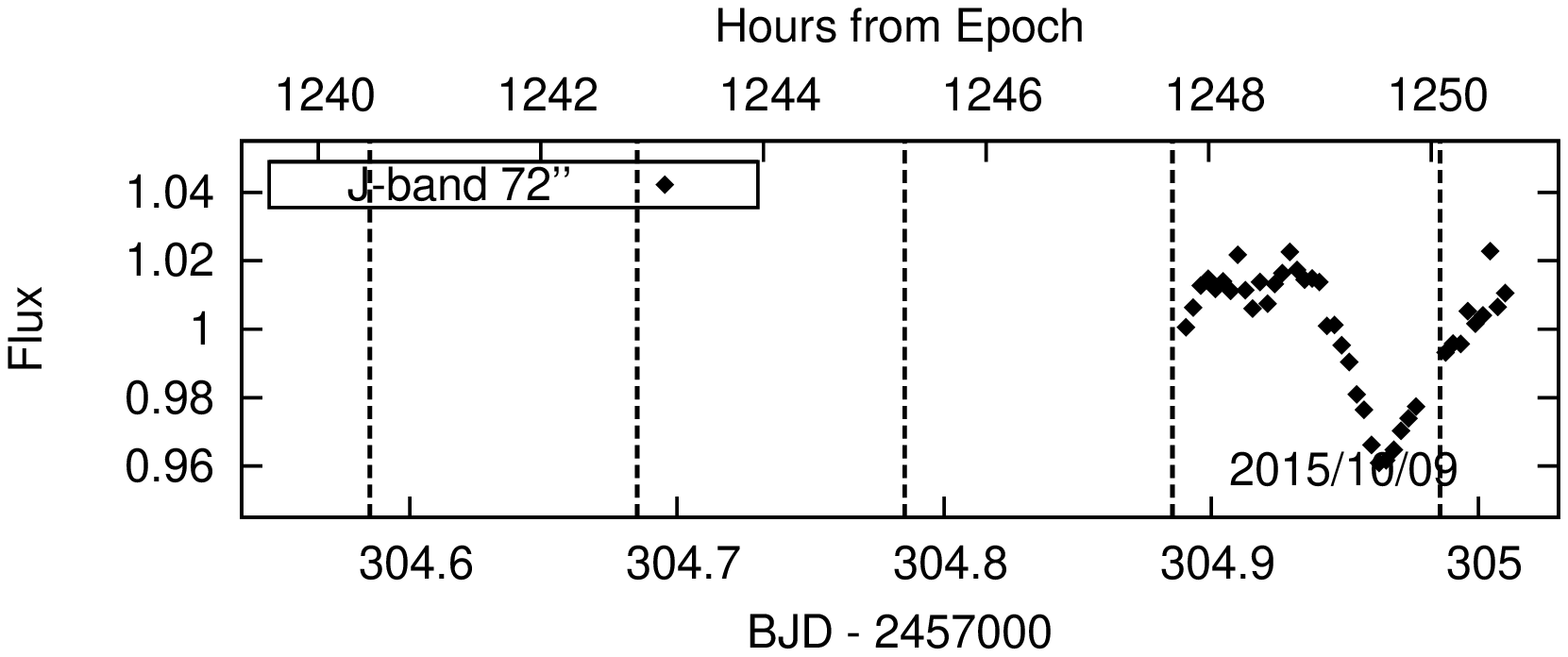}
\includegraphics[scale=0.50, angle = 270]{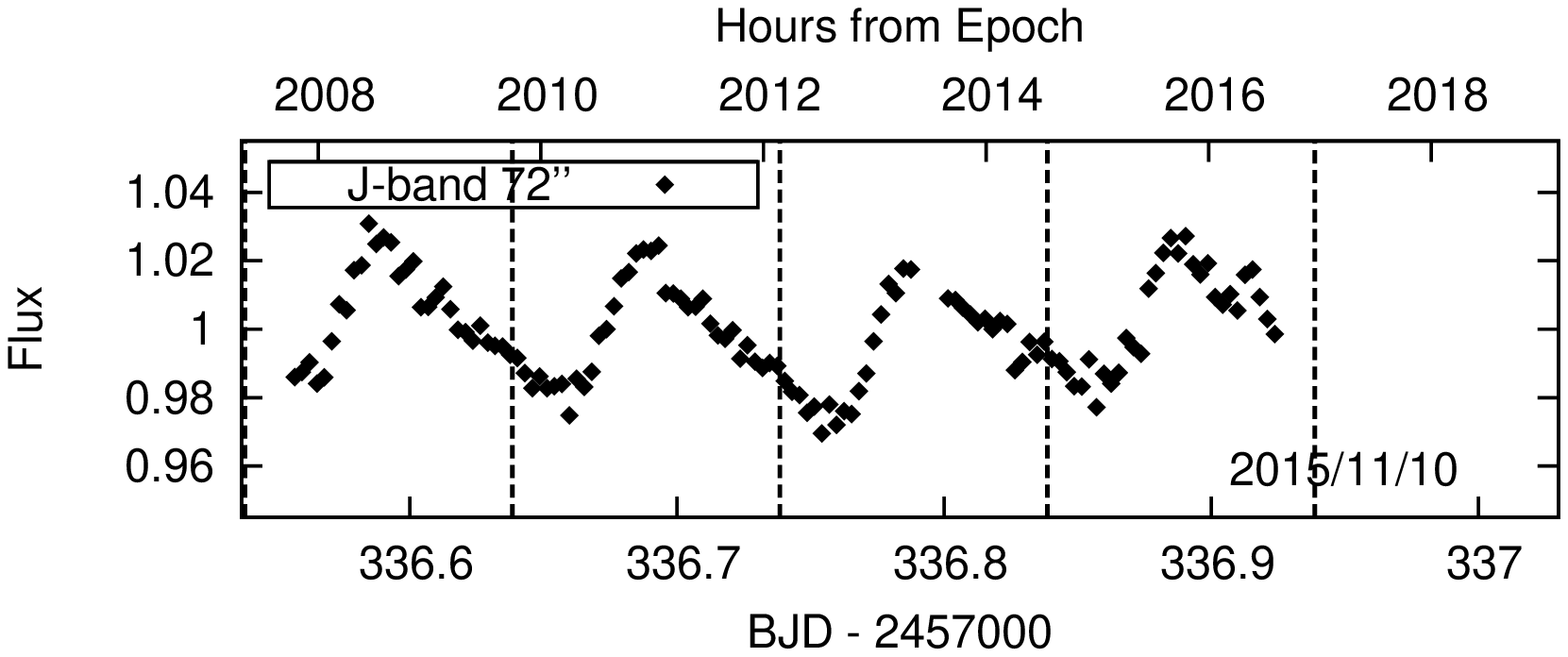}
\includegraphics[scale=0.50, angle = 270]{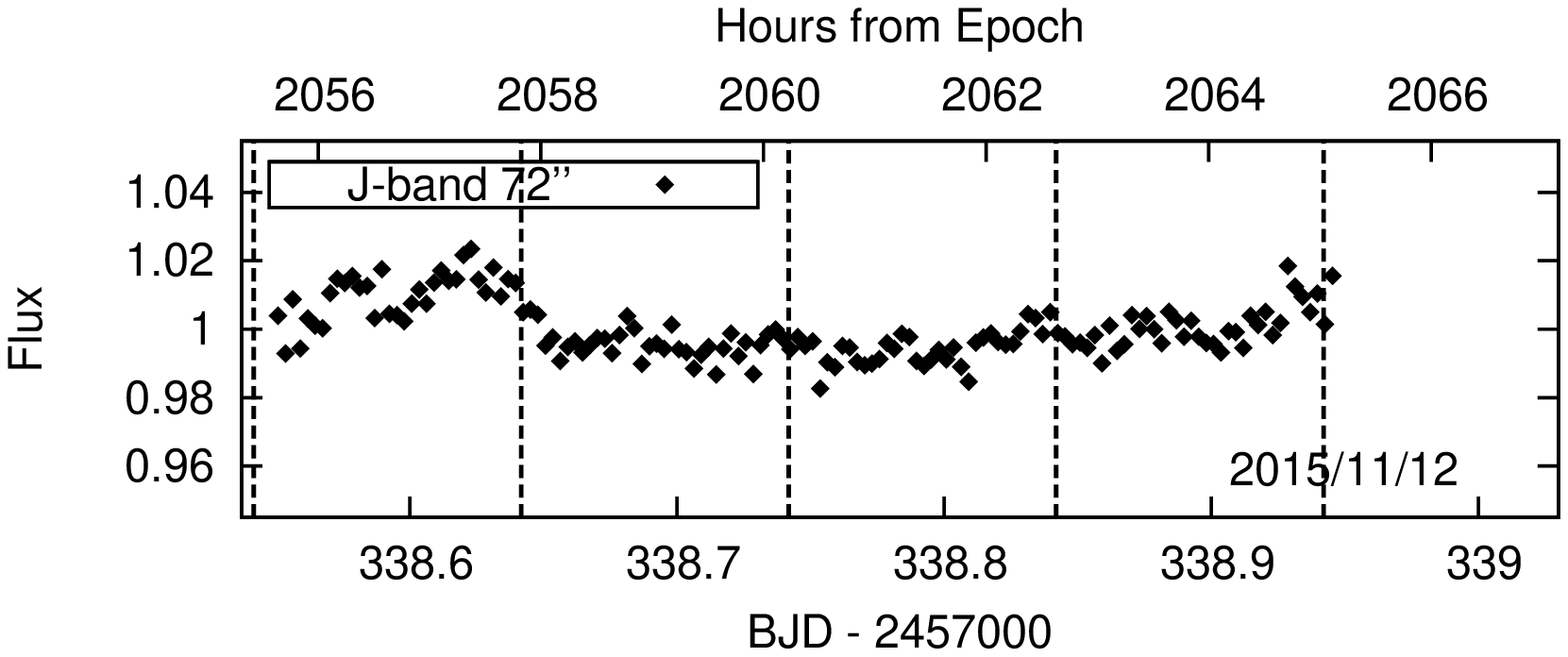}
\includegraphics[scale=0.50, angle = 270]{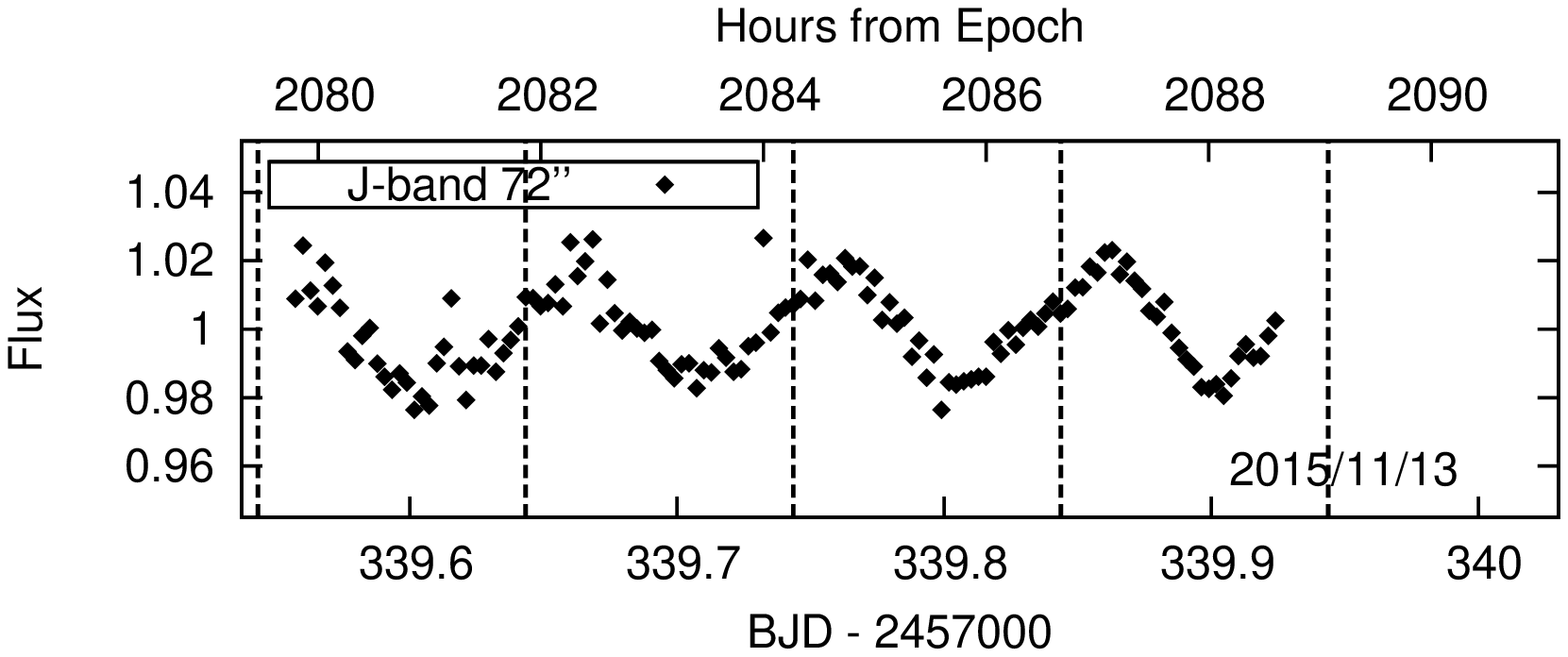}
\includegraphics[scale=0.50, angle = 270]{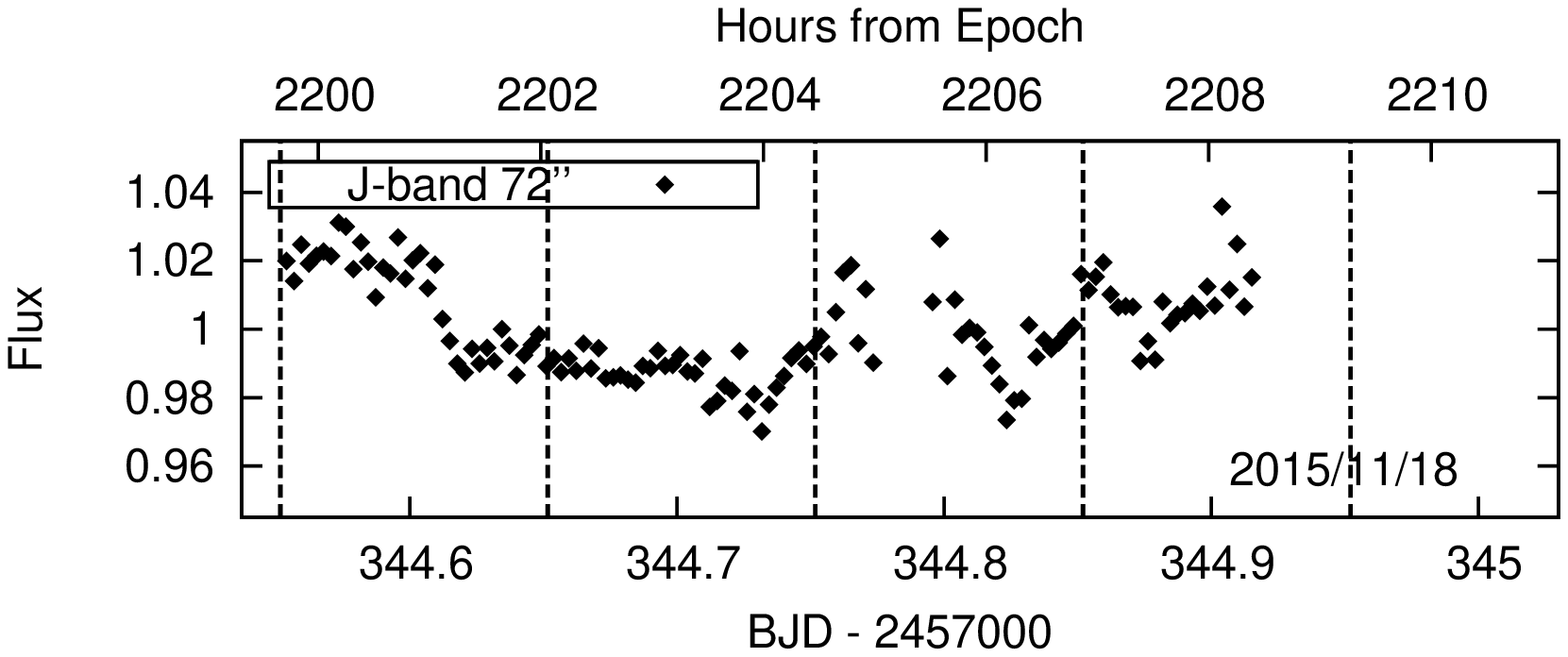}
\caption[]
	{	
		%J-band photometry of SIMP 0136
		Photometry of SIMP 0136 on the dates given in the lower-right of each panel (UT)
		with the Perkins 1.8-m (72'') telescope,
		at various wavelengths as indicated in the panels.
		The vertical dashed lines indicate 
		cycles of the apparent $\sim$\WeightedMeanSIMPPeriod \ hour rotation period of SIMP 0136, compared
		to the apparent flux maximum in our J-band photometry at a barycentric julian date (BJD) of: BJD-2457000 $\sim$252.899.
		For clarity we plot our photometry binned every $\sim$4 minutes ($\sim$0.003 $d$).
	}
\label{FigSIMP0136One}
\end{figure*}

\begin{figure*}
% \centering
\includegraphics[scale=0.50, angle = 270]{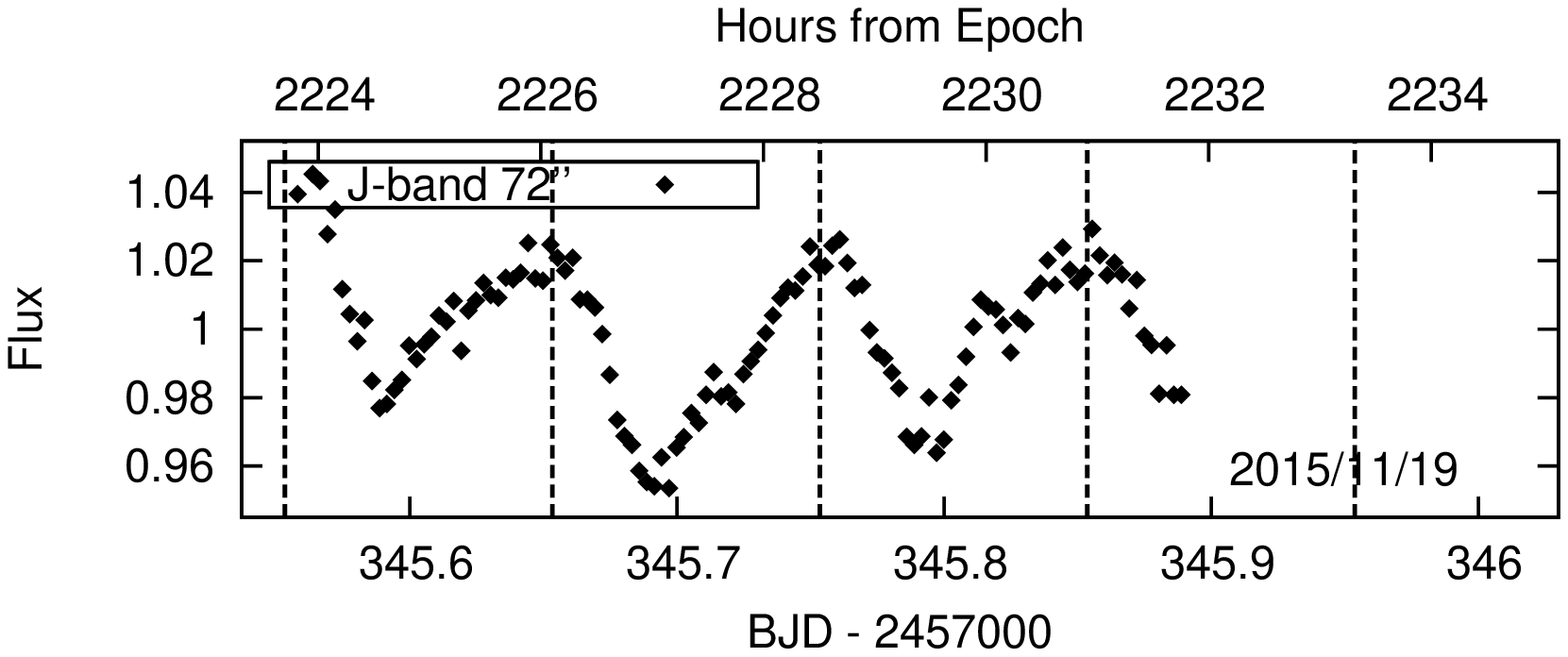}
\includegraphics[scale=0.50, angle = 270]{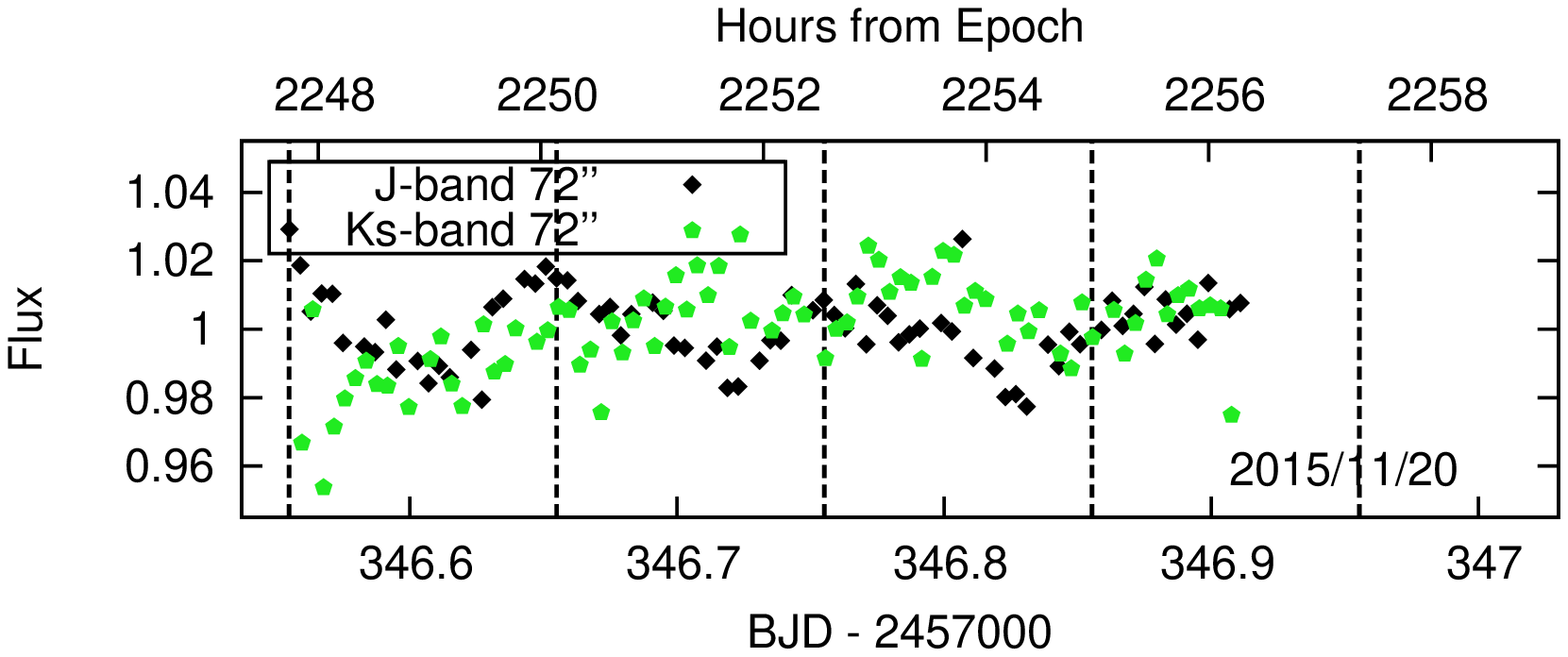}
\includegraphics[scale=0.50, angle = 270]{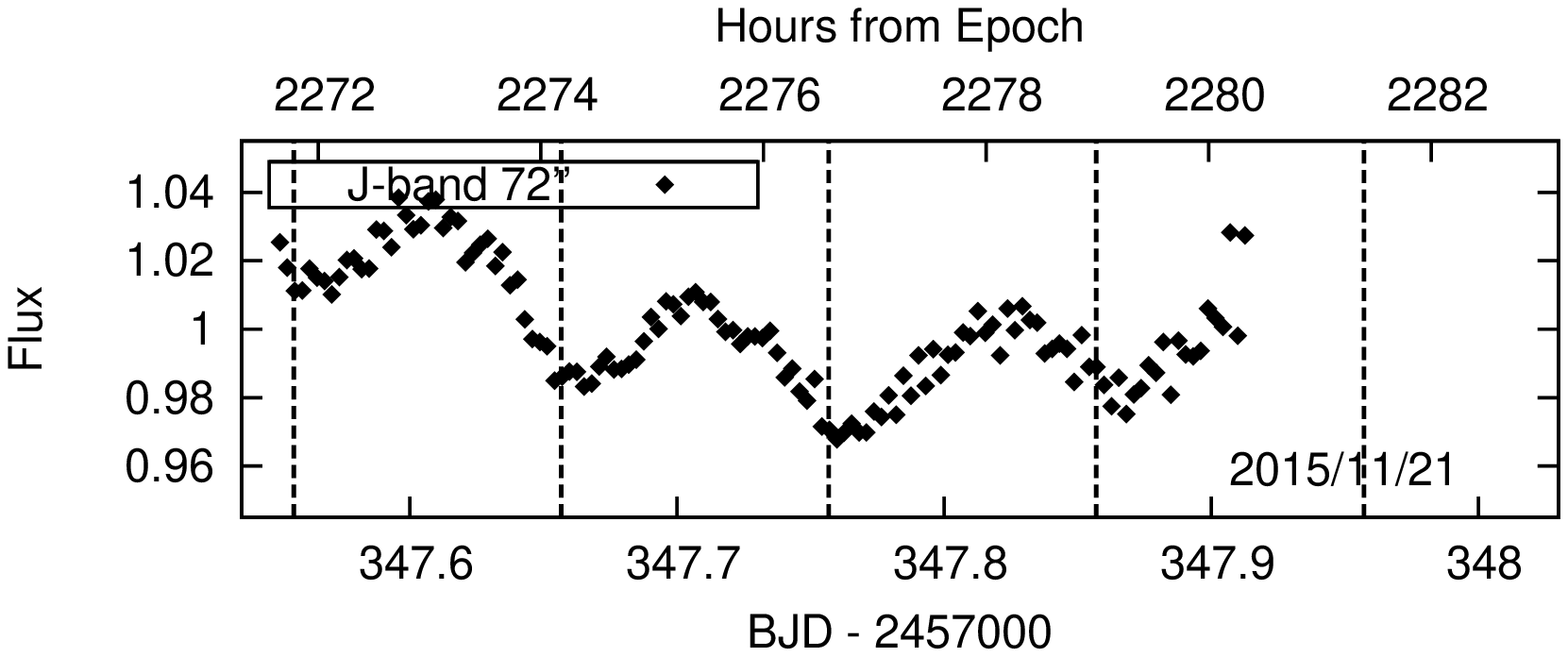}
\includegraphics[scale=0.50, angle = 270]{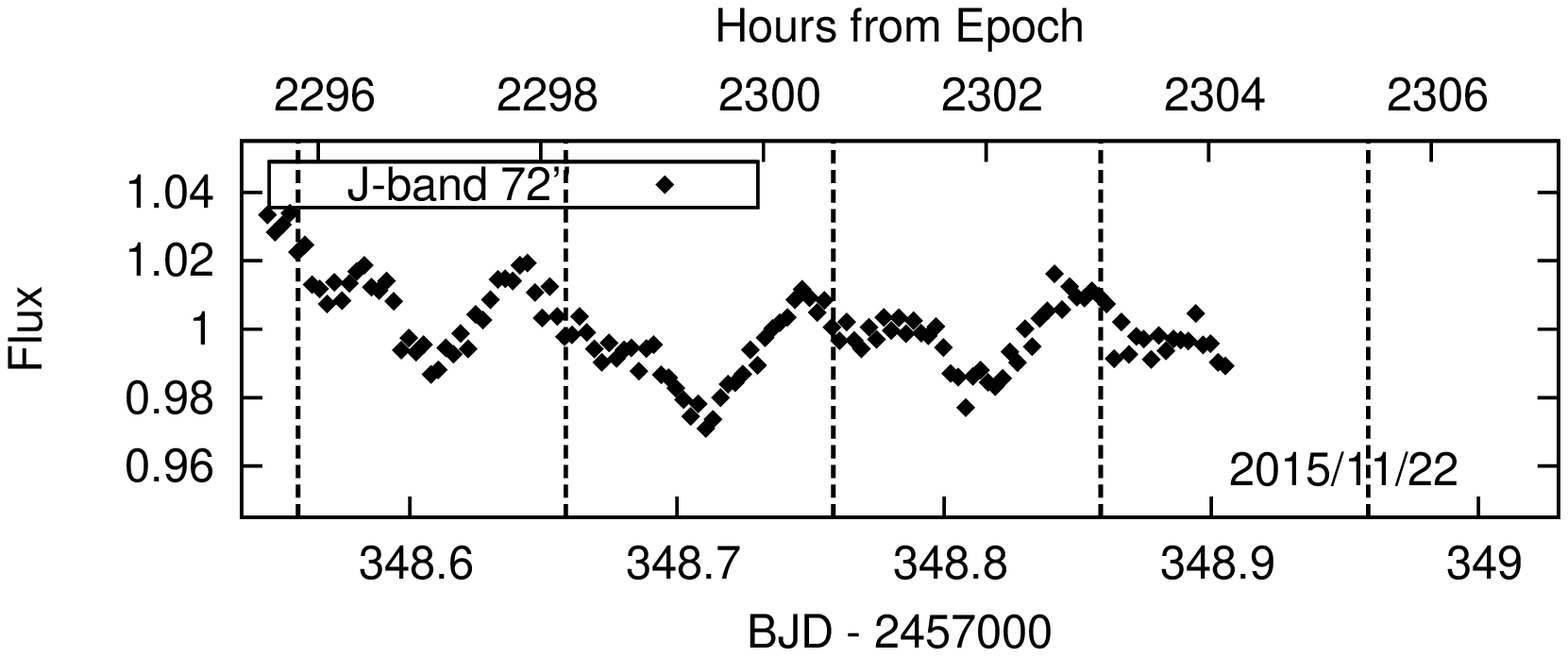}
\includegraphics[scale=0.50, angle = 270]{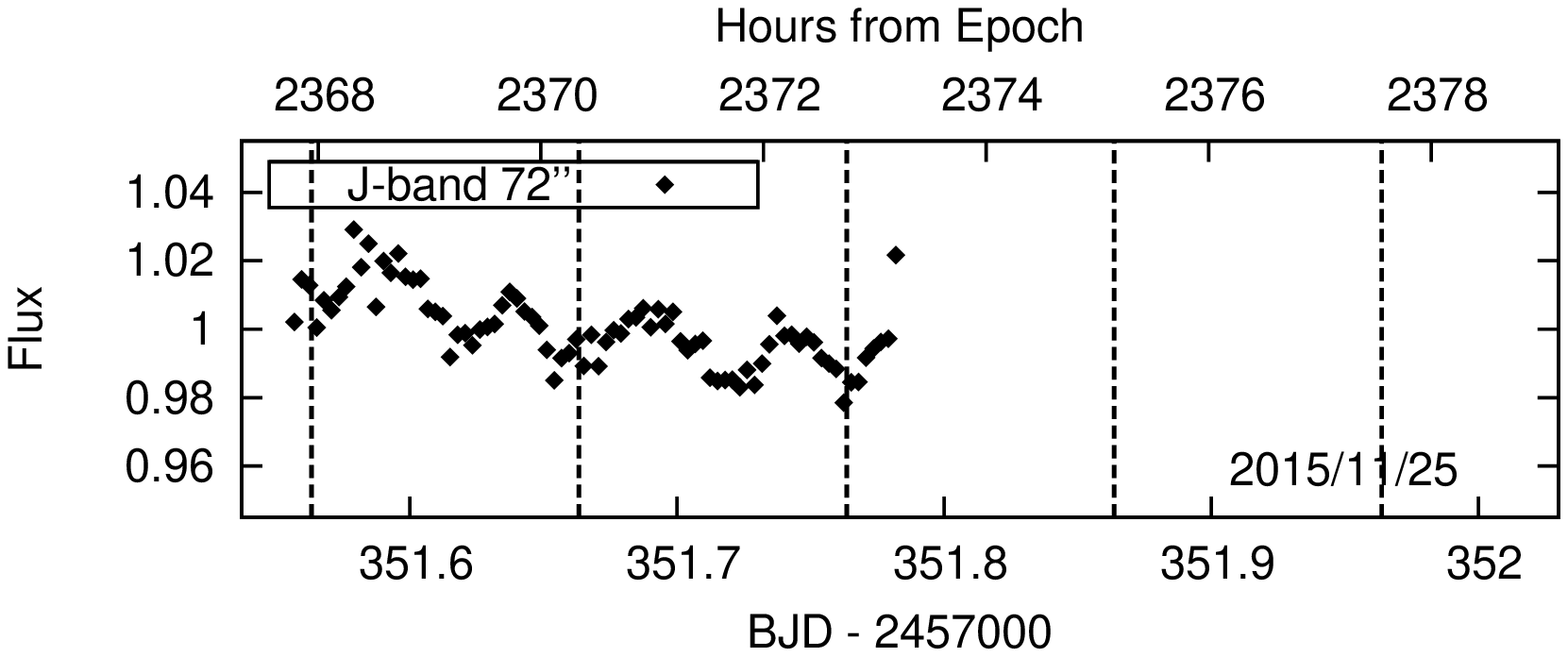}
\includegraphics[scale=0.50, angle = 270]{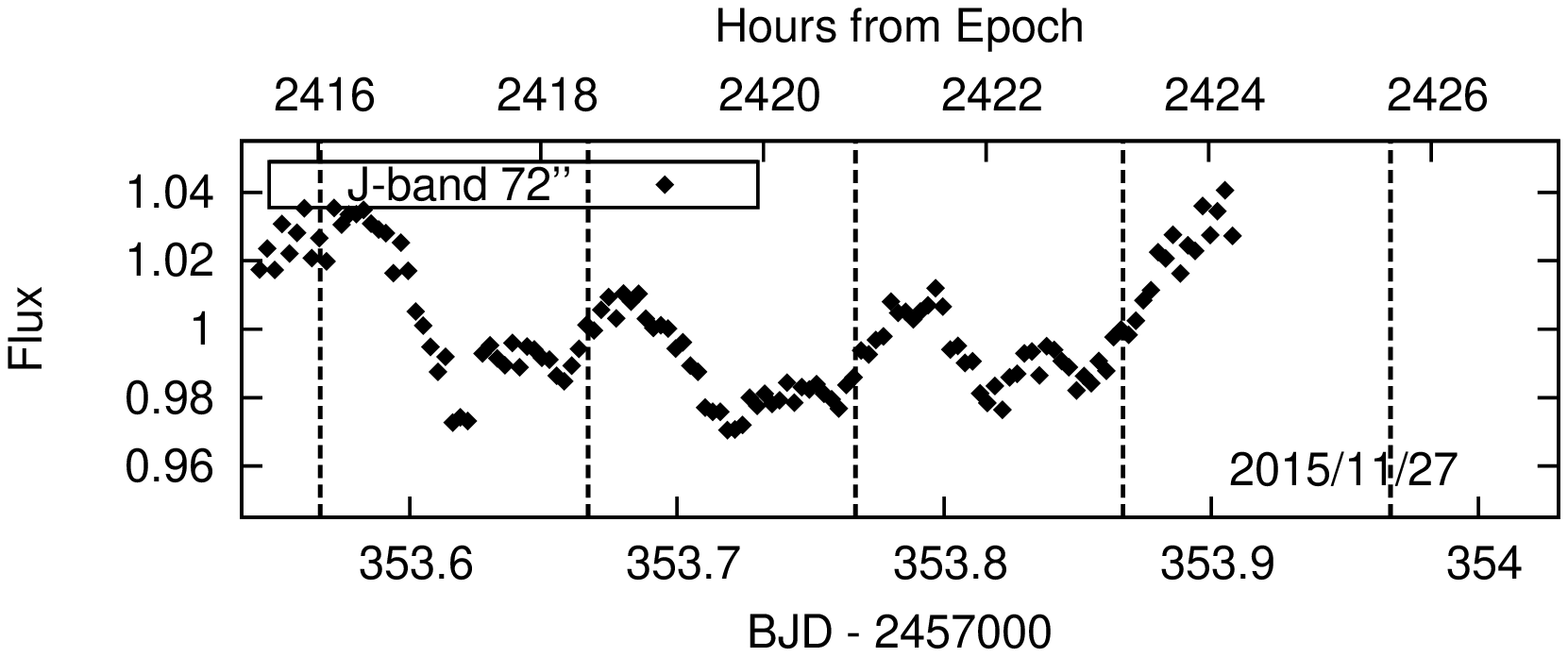}
\includegraphics[scale=0.50, angle = 270]{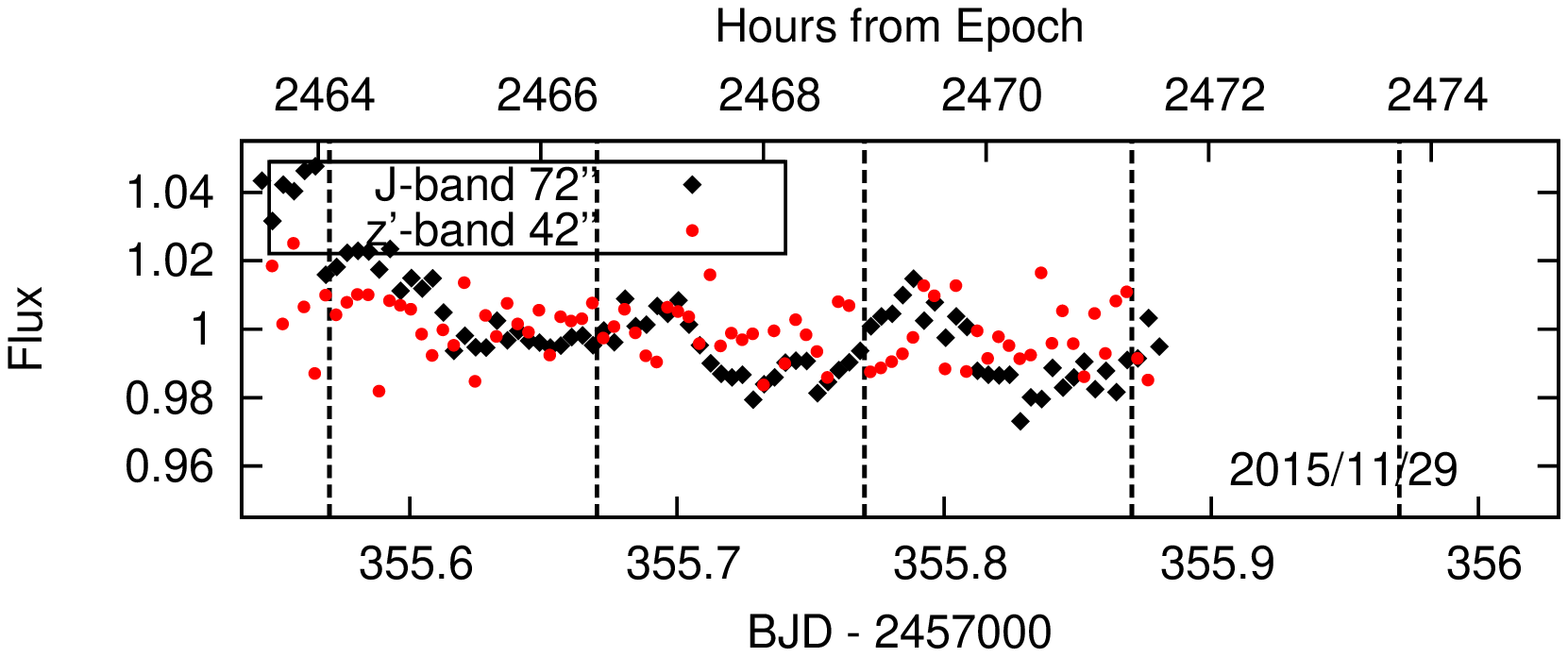}
\includegraphics[scale=0.50, angle = 270]{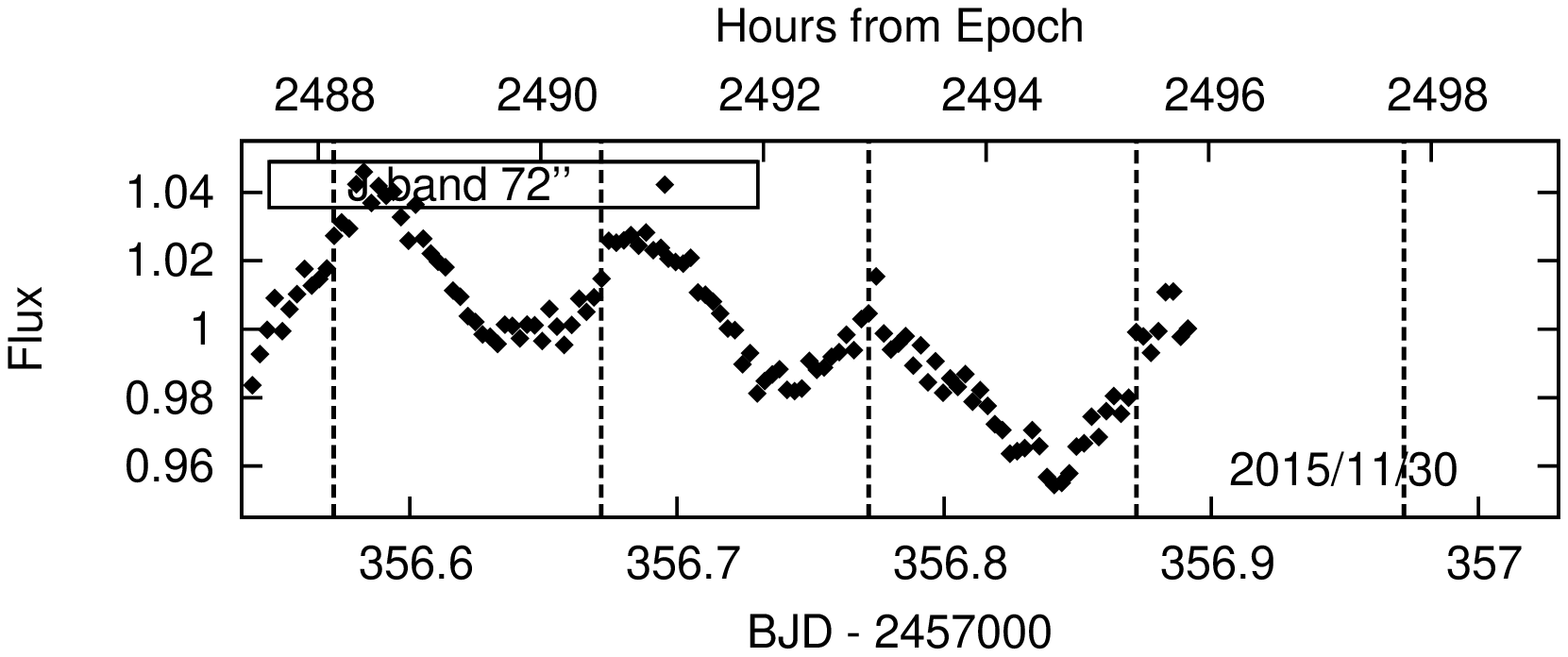}
\includegraphics[scale=0.50, angle = 270]{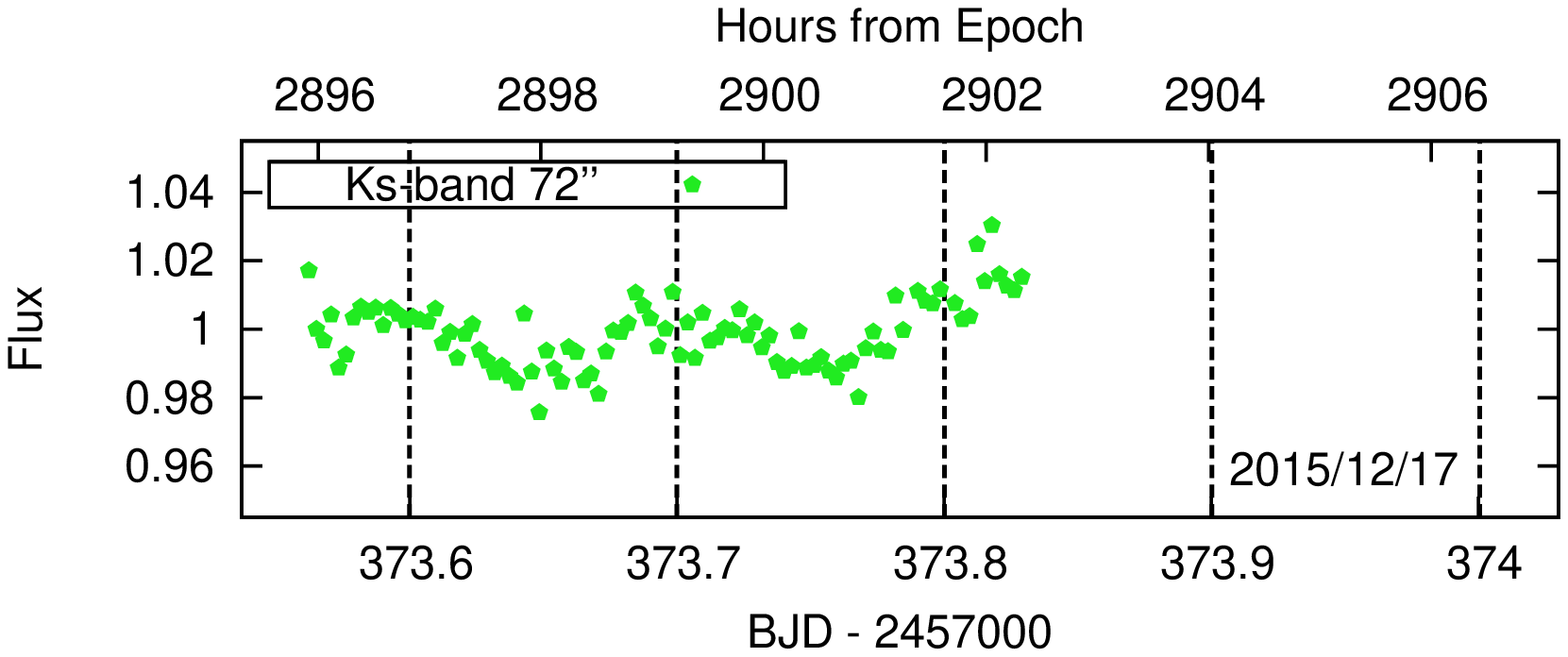}
\caption[]
	{	Photometry of SIMP 0136 on the dates given in the lower-right of each panel (UT)
		with the Perkins 1.8-m (72'') and Hall 1.1-m (42'') telescopes,
		at various wavelengths as indicated in the panels.
		The vertical dashed lines indicate 
		cycles of the apparent $\sim$\WeightedMeanSIMPPeriod \ hour rotation period of SIMP 0136, compared
		to the apparent flux maximum in our J-band photometry at BJD-2457000 $\sim$252.899.
		For clarity we plot our photometry binned every $\sim$4 minutes ($\sim$0.003 $d$)
		for panels featuring single
		wavelength photometry, and binned every 
		$\sim$6 minutes (0.004 $d$) for panels featuring multiwavelength photometry.

	}
\label{FigSIMP0136Two}
\end{figure*}

%Still some DCT data that I can analyze...
\begin{figure}
% \centering
\includegraphics[scale=0.50, angle = 270]{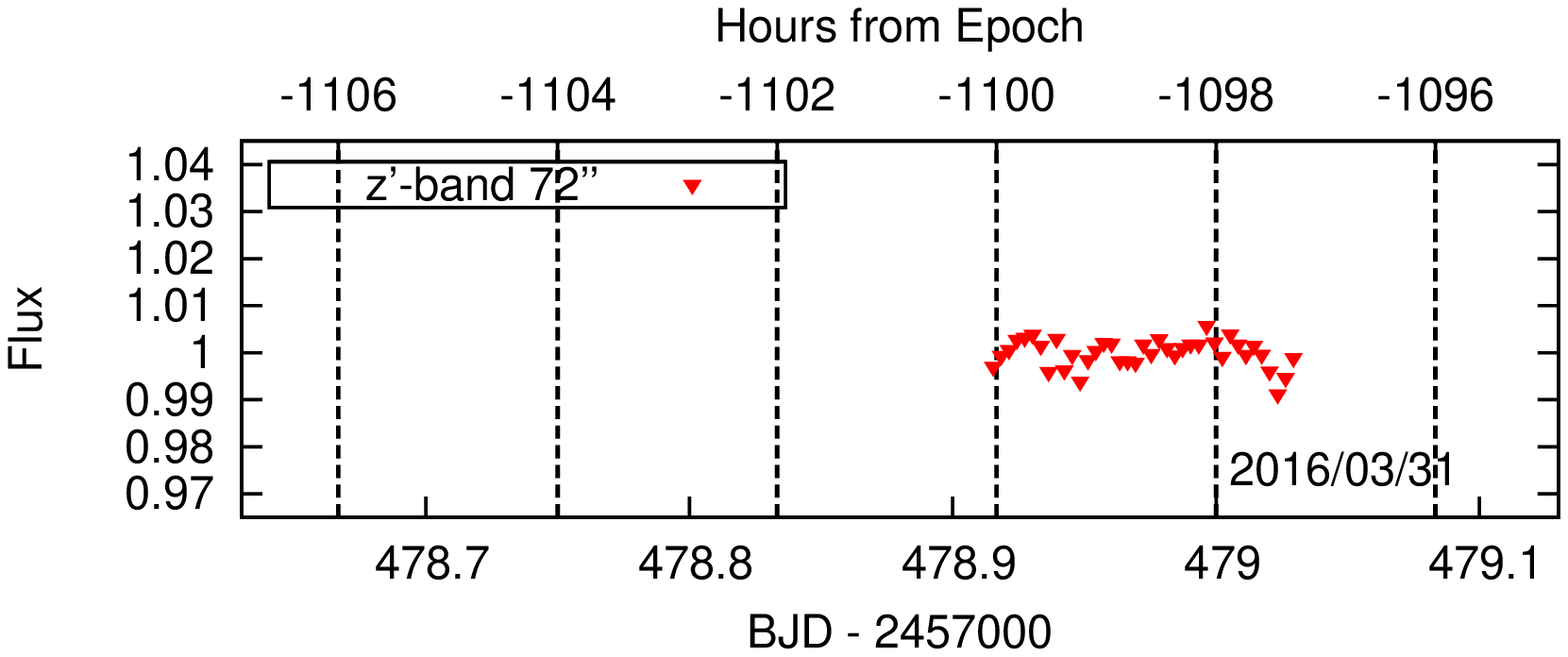}
\includegraphics[scale=0.50, angle = 270]{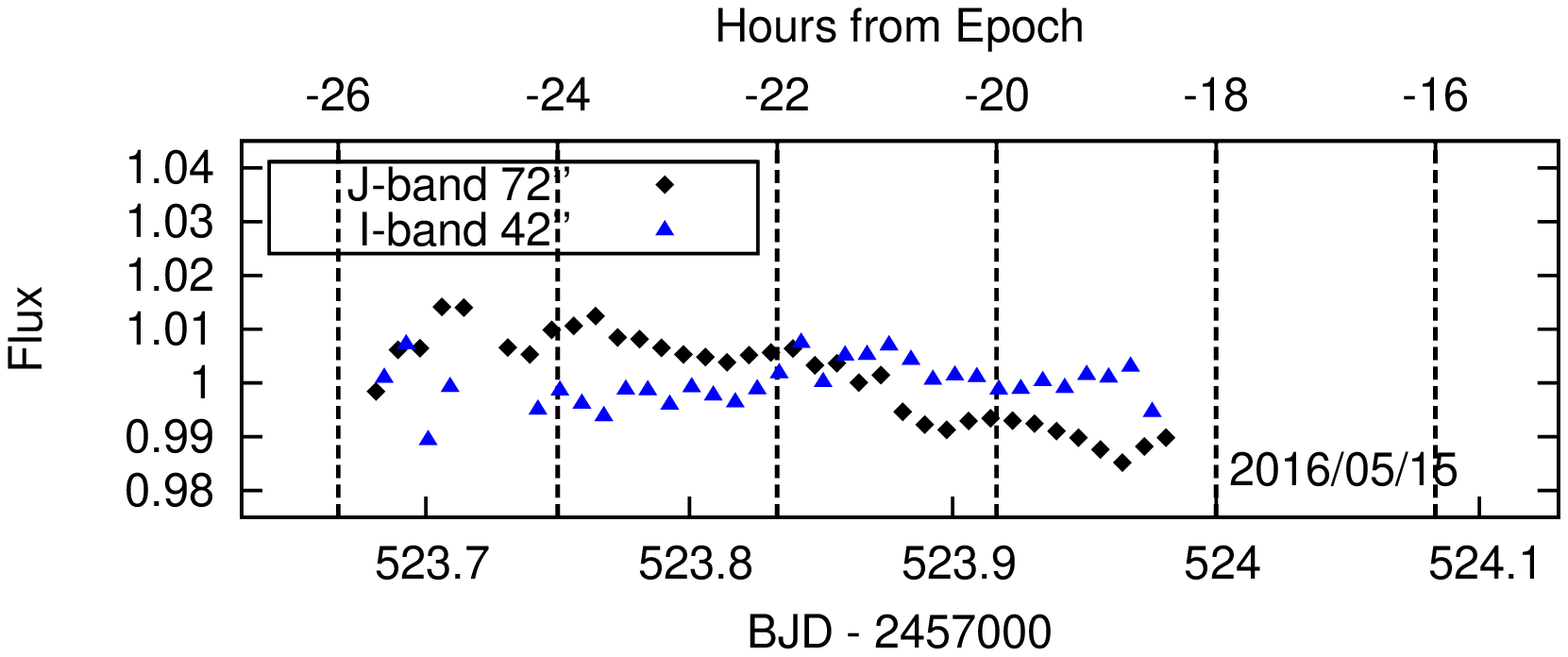}
\includegraphics[scale=0.50, angle = 270]{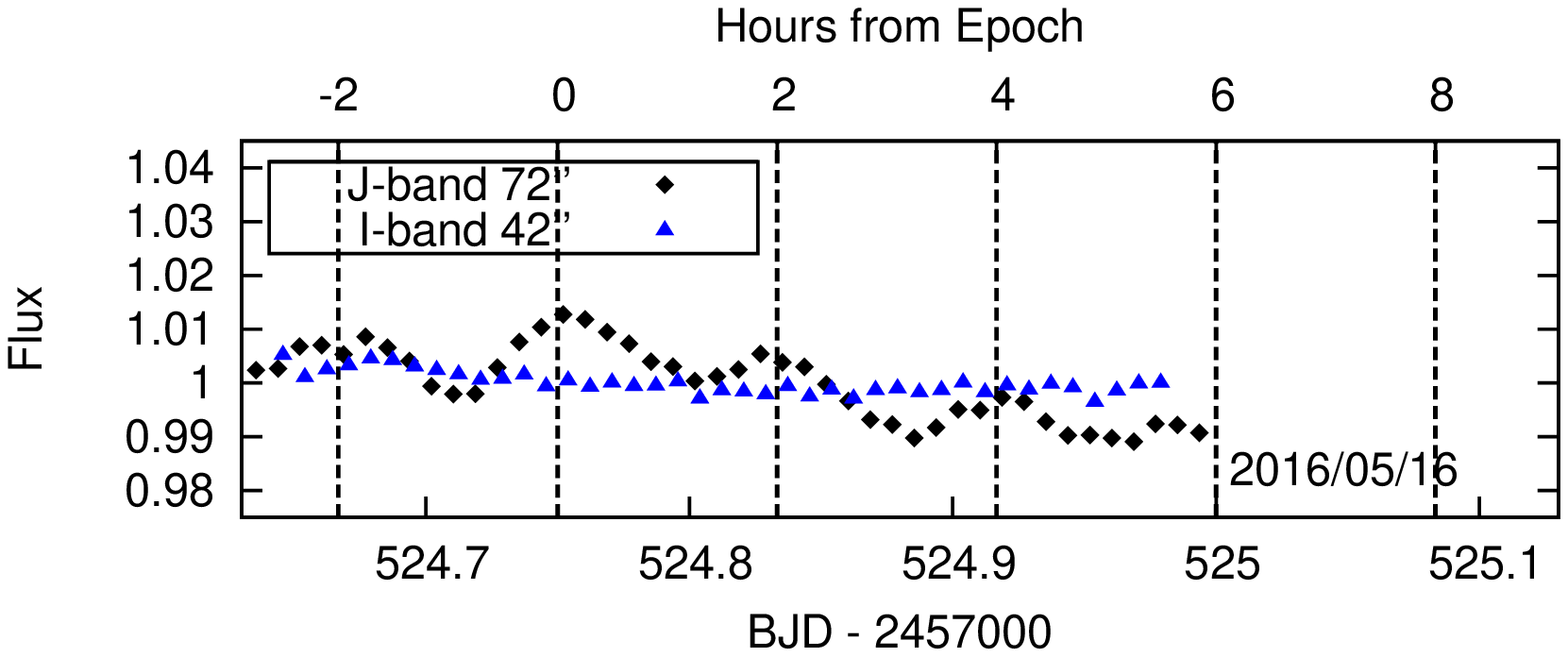}
\caption[]
	{	Photometry of TVLM 513 on the dates given in the lower-right of each panel (UT)
		in various bands as indicated in the panels.
		The vertical dashed lines indicate 
		cycles of the apparent $\sim$2.0 hour rotation period of TVLM 513-46546 \citep{Hallinan07,Harding13},
		compared
		to the apparent flux maximum in our J-band photometry at BJD-2457000 $\sim$524.75.
	}
\label{FigTVLM}
\end{figure}

 We observed the ultra-cool dwarfs SIMP 0136 and TVLM 513 using a variety of telescopes
at the Lowell Observatory. 
SIMP 0136 was observed for \INSERTNIGHTSSIMP \ nights spread out over \INSERTSPREADDAYSSIMP\ days,
while 
TVLM 513 was observed for \INSERTNIGHTSTVLM \ nights over \INSERTSPREADDAYSTVLM\ days.
The observations of these dwarfs were conducted using the Perkins 1.8-m telescope, and the Hall 1.1-m telescope.
%%%%%%%and the 4.3-m Discovery Channel Telescope (DCT). 
Our photometry from the Perkins telescope was obtained using either the PRISM imager \citep{Janes04}
in the optical and very near-infrared,
or the Mimir instrument \citep{Clemens07} in the near-infrared.
% On the DCT we used the Large Monolithic
% Imager (LMI; Massey et al. 2013), and on the Hall telescope we used the NASA42
% imager.
On the Hall telescope we used the NASA42 imager.

Our optical and near-infrared photometry was reduced and analyzed using the techniques described fully in \citet{CrollTwoMass}.
Unlike the data presented in \citet{CrollTwoMass} we do not attempt to employ fringe frames to remove the obvious fringes in any of our I-band or
z'-band data on the Hall or Perkins telescope.
For our Perkins/Mimir data we employ the \citet{Clemens07} non-linearity correction;
we have also repeated the forthcoming analysis not using a non-linearity correction for our Perkins/Mimir data 
and note that the major conclusions of the paper are basically unchanged.
We summarize our SIMP 0136 observations in Table \ref{TableSIMPObs} and our TVLM 513 observations in Table \ref{TableTVLMObs}.
All dates given in this paper are UT dates.
In these tables we give the aperture sizes, and annuli sizes, we used for our aperture photometry. 
Our SIMP 0136 light curves are displayed in Figure \ref{FigSIMP0136One} and \ref{FigSIMP0136Two},
while our TVLM light curves are displayed in Figure \ref{FigTVLM}.

% \section{Analysis \& Discussion}
% \label{SecAnalysis}

\section{SIMP 0136}
\label{SecAnalysisSIMP}

\subsection{The Rotation Period of SIMP 0136}

\begin{figure}
% \centering
\includegraphics[scale=0.45, angle = 270]{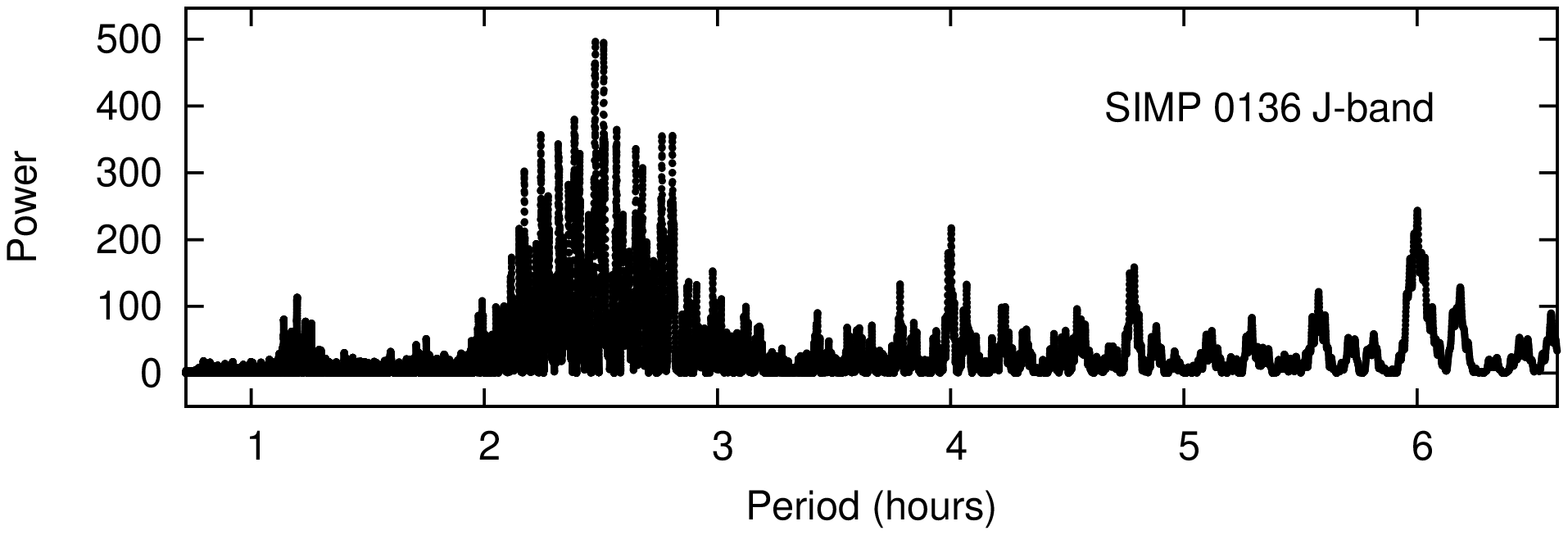}
\includegraphics[scale=0.45, angle = 270]{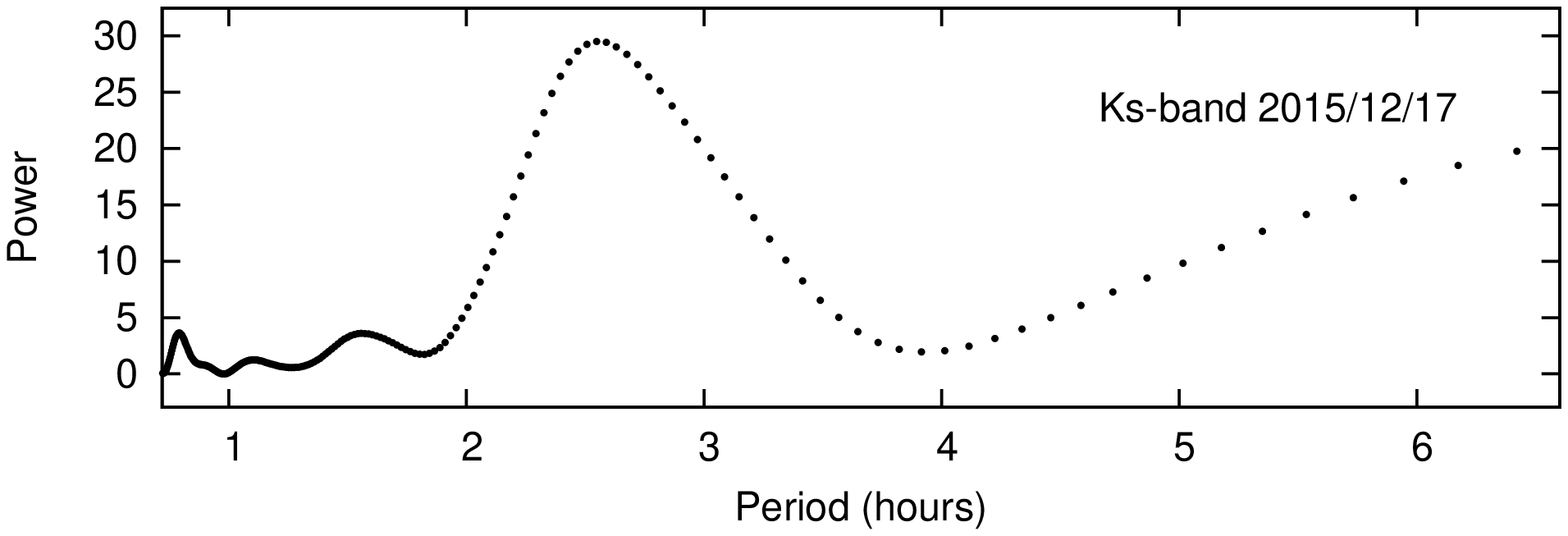}
\caption[]
	{	Lomb-Scargle Periodogram of our SIMP 0136 photometry in J-band (top), 
		and in Ks-band on 2015/12/17 (bottom; note the different vertical scales on the y-axis).
		The peak periodogram value in our J-band photometry is
		\PeriodHoursSIMPZeroOneThreeSixJband \ $\pm$ \PeriodHoursErrorSIMPZeroOneThreeSixJband \ hours,
		while for our Ks-band photometry on 2015/12/17 the peak periodogram value is
		\PeriodHoursSIMPZeroOneThreeSixKsbandFifteenDecemberSeventeen \ $\pm$ \PeriodHoursErrorSIMPZeroOneThreeSixKsbandFifteenDecemberSeventeen \ hours.
		%and with all our photometry (bottom). We present the peak periodogram values in Table \ref{TablePeriods},
		%which are all approximately $\sim$\PeriodHoursTwoMassZeroZeroThreeSixApprox \ hours.
		
	%Lomb-Scargle Periodograms of our photometry of SIMP 0136 in various bands as indicated in the panels, 
	%and with all our photometry (bottom). We present the peak periodogram values in Table \ref{TablePeriods},
	%which are all approximately $\sim$\PeriodHoursTwoMassZeroZeroThreeSixApprox \ hours.
	}
\label{FigSIMP0136LombPeriodogram}
\end{figure}

\begin{figure*}
% \centering
\includegraphics[scale=1.07, angle = 270]{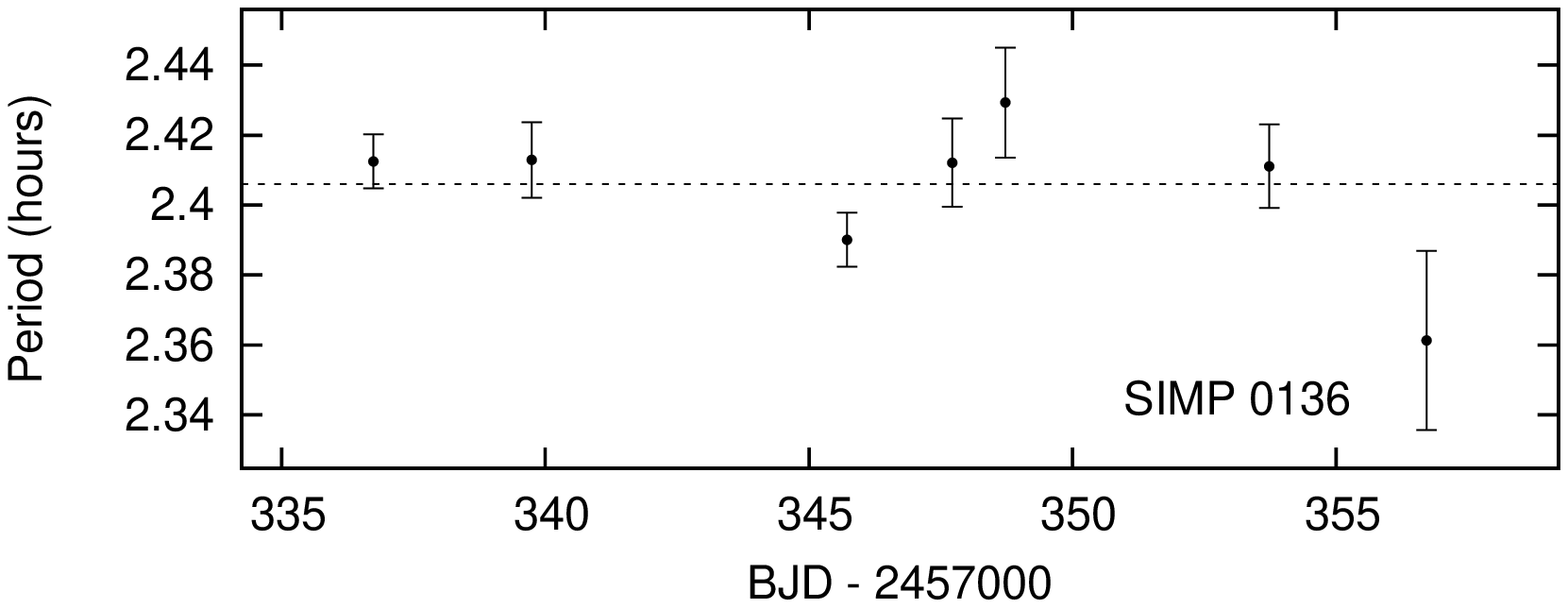}
\caption[]
	{	Photometric starspot model estimates of the photometric
		period of SIMP 0136 
		from our highest precision, J-band light curves with the least light curve evolution 
		(2015 November 10, 	
		2015 November 13, 
		2015 November 19, 
		2015 November 21, 
		2015 November 22, 
		2015 November 27, 
		\& 2015 November 30.)
		The dashed horizontal line represents the weighted mean of these various estimates of the 
		rotation period.
		There is no strong evidence for variations in the estimates
		of the cloud rotation periods from night-to-night, and 
		therefore we do not present evidence in favour of differential rotation, or similar phenomena.
		If differential rotation dominated periodicity exists during an entire night of our observations,
		it must cause variations in the photometric 
		period of less than $\sim$\DiffRotationPercentage\%.
	}
\label{FigSIMP0136PeriodEvolution}
\end{figure*}

Here we attempt to determine the rotation period of SIMP 0136 from our photometric light curves.
The quasi-sinusoidal variability displayed in our SIMP 0136 light curves suggest that 
Lomb-Scargle periodograms (\citealt{Lomb76}; \citealt{Scargle82}) may be a suitable
method to determine the rotation period of SIMP 0136.
We present Lomb Scargle periodograms of all our J-band photometry of SIMP 0136 in Figure \ref{FigSIMP0136LombPeriodogram}.
We utilize the technique discussed in \citet{CrollTwoMass} to estimate the associated errors on the peak periodogram value, and
infer
a photometric
period of \PeriodHoursSIMPZeroOneThreeSixJband \ $\pm$ \PeriodHoursErrorSIMPZeroOneThreeSixJband \ hours.
Similarly, we present a Lomb-Scargle periodogram on our highest quality
Ks-band light curve of SIMP 0136 (Figure \ref{FigSIMP0136LombPeriodogram});
the peak periodogram value from our 2015 December 17 Ks-band photometry of SIMP 0136 is
\PeriodHoursSIMPZeroOneThreeSixKsbandFifteenDecemberSeventeen \ $\pm$ \PeriodHoursErrorSIMPZeroOneThreeSixKsbandFifteenDecemberSeventeen \ hours.
Both these values are slightly longer than a previous estimate of the rotation period of SIMP 0136;
\citet{Artigau09} estimated the rotation period of SIMP 0136 as 
2.3895 $\pm$ 0.0005 hours
from fitting two sinusoids to their two nights of double-peaked quasi-sinusoidal J-band photometry.

We note that the periodic repetition of features, such as maxima or minima,
in the light curve of SIMP 0136 may not provide an accurate estimate of the rotation period
of this ultra-cool dwarf.
The first reason this might be the case, is due to evolution of the SIMP 0136 light curve.
Our SIMP 0136 light curves show occasion significant evolution in the apparent longitude of minima \& maxima,
compared to the nominal
rotation period of this L/T transition brown dwarf.
On the night of 2015 November 12 the variability
of SIMP 0136 nearly completely disappears,
with peak-to-peak amplitudes of less than $\sim$0.5\%. 
The phase of the maxima from two days prior (2015 November 10) to a
day after (2015 November 13) this date, 
appear to flip, suggesting that the longitude of the cloud 
gaps\footnote{For the following discussion we assume that the SIMP 0136 variability is driven by thinner
gaps in thick clouds \citep{Artigau09,Apai13}, but our conclusions are essentially unchanged if the variability
is driven by spots, aurorae, different cloud behaviour, or another mechanism.}
leading to the observed variability
have significantly changed.
This large change in longitude of the apparent cloud gaps may have been caused by the cloud
gaps completely disappearing and reappearing, or a gradual evolution in the longitude of these cloud gaps (due to
spot evolution or differential rotation, for instance).
If the longitudes of these gaps in clouds are constantly evolving from night-to-night
then a Lomb-Scargle periodogram of a long-term data-set 
may not give the best estimate of the rotation period of this brown dwarf.

Another method that has been used previously to estimate the rotation period 
of SIMP 0136 from photometric light curves \citep{Artigau09} has been to fit a short segment of data
that appears to display obvious periodicity and limited light curve evolution.
As long as the longitude of cloud gaps on SIMP 0136 are stable for a few rotation periods
at a time then this method should give a precise estimate of the photometric period.
Our highest precision light curves that display the most obvious periodicity with little evolution during
the night are the following light curves:
our J-band light curves on
2015 November 10, 
2015 November 13, 
2015 November 19, 
2015 November 21, 
2015 November 22, 
2015 November 27, 
\& 2015 November 30.
To determine the photometric period from these nights of photometry we fit the light curves
with circular spots employing
a \citet{Budding77} model
utilizing the methodology discussed in {\it StarSpotz} \citep{Croll06,CrollMCMC,Walker07}.
Although photometric starspot models are notoriously non-unique (e.g. \citealt{CrollMCMC,Walker07,Aigrain15}),
in general they are well-suited to recover the period of variability \citep{Aigrain15}.
We therefore fit each of our light curves with a two starspot model utilizing the Markov Chain Monte
Carlo functionality we discuss in \citet{CrollMCMC};
these starspot model fits are not intended to be unique, but only to return an accurate estimate of the period.
For the following nights we remove
linear trends from the light curves (see Section \ref{SecVariabilityNight})
to ensure the light curve maxima display relatively constant flux: 
2015 November 21,
2015 November 22,
\& 2015 November 30.
We increase our photometric error bars during intervals that we believe the data might be affected by systematic errors (such
as due to passing clouds, or just before/after civil twilight);
we subsequently also scale our photometric uncertainties to ensure
the reduced $\chi^2$ of the resulting {\it StarSpotz} fits are approximately one.
We inspect each one of the resulting fits from the individual nights to ensure 
the resulting spot model period accurately tracks both the minima and maxima in the
light curves. 
All our light curves appear to be well fit by a single period, except our 2015 November 30 light curve.
A single photometric period does not appear to track all the minima and maxima of our 2015 November 30 light curve,
and therefore we artificially scale up the errors on the period determination from that night by a
factor of \SIMPPeriodScaleFactorFifteenNovemberThirtyFactor.
We display the photometric period estimates from the individual nights in
Figure \ref{FigSIMP0136PeriodEvolution}.
The weighted mean and error of these estimates of the photometric period 
are
\WeightedMeanSIMPPeriod \ $\pm$ \WeightedErrorSIMPPeriod \ hours.

Given the obvious evolution in variability that we observe,
we suggest that the combination of the independent estimates 
of the photometric period of SIMP 0136,
from our best nights of photometry (those with the most obvious periodicity, and displaying little light curve evolution),
provides our most suitable estimate of its photometric period.
As we occasionally observe light curve evolution even within a single night of observations,
but our spot model assumes constant spots for each night of observations, we artificially
scale up the uncertainty on our period estimate to take this evolution into account;
we choose to scale up the error by a factor of \SIMPPeriodScaleFactor.
Therefore, we quote
the photometric period, and therefore the likely rotation period, of the L/T transition dwarf SIMP 0136
as
\WeightedMeanSIMPPeriod \ $\pm$ \WeightedErrorSIMPPeriodScaled \ hours.

%; the standard deviation is  \ hours

% The reduced $\chi^2$ of these estimates of the cloud rotation period on different nights compared
% to this weighted mean is nearly exactly 1, and therefore there is no strong evidence for variability
% in the cloud rotation period that might result from differential rotation or a similar mechanism.

% % We first perform Lomb-Scargle periodograms on each of these individual light curves, and display
% % the peak periodogram values and related errors in Figure \ref{FigSIMP0136PeriodEvolution}.

These independent estimates of the photometric period of SIMP 0136 from night-to-night,
can also be used to search for variations in the photometric period,
such as could be caused by differential rotation with latitude,
or different wind speeds at different epochs, or at different depths in the atmosphere.
Jupiter experiences differential rotation, 
as clouds near Jupiter's pole rotate 5 minutes slower
than the equator \citep{NewburnGulkis73}.
SIMP 0136 may display similar behaviour.
To determine if 
the photometric period of our SIMP 0136 light curves evolves from night-to-night
we utilize the same high precision light curves without significant light curve evolution, and the associated
{\it StarSpotz} fits (Figure \ref{FigSIMP0136PeriodEvolution}).
The reduced $\chi^2$ of these estimates of the cloud rotation period on different nights compared
to this weighted mean is $\sim$\ReducedChiSquaredDiffRotation. 
Given that we argued that it was justified above to scale up the errors on our period estimates
(to take into account that in some cases 
we observe light curve evolution within a single night of observations, but our spot model assumes constant spots),
if we do likewise here then the reduced $\chi^2$  
\DiffRotationReducedChiSquaredComment. 
%If we scale up the errors on these period estimates from 
%separate nights by a factor of \SIMPPeriodScaleFactor \ (as we argue is justified for the weighted error above to take into account that in some cases 
%we observe light curve evolution within a single night of observations, but our spot model assumes constant spots),
%then the reduced $\chi^2$ is 
%\DiffRotationReducedChiSquaredComment. 
Therefore, we believe there is no strong evidence for variability
in the photometric period from night-to-night that might result from differential rotation or a similar mechanism.
We note that within a single night of photometry, we find no obvious evidence for two different spot/cloud groups
rotating with different periods and therefore differentially, as has been previously obviously observed for solar
analogues (e.g. \citealt{Rucinski04,Walker07}).
If differential rotation is occurring during our six nights of precise photometry with little light curve evolution,
and this differential rotation results in a different photometric period for a complete night of our observations,
then the range in periods must be less than
$\sim$\DiffRotationPercentage\% (as calculated by comparing the period of the most discrepant night to our weighted mean).

We also perform a {\it StarSpotz} fit to our Ks-band light curve on 
2015 December 17 that similarly displays little light curve evolution.
The resulting period is
\PeriodHoursFifteenDecemberSeventeenKsBand \ $\pm$ \PeriodHoursErrorFifteenDecemberSeventeenKsBand \ hours.
% Our 2015 December 17 Ks-band light curve is not of sufficient duration, or accuracy...
As our Ks-band photometric period from our 2015 December 17 light curve
is not sufficiently different from our estimate of the J-band photometric period,
we find no evidence for different rotation periods 
at the arguably different depths in the atmosphere probed by our Ks-band observations.
Our z'-band light curves, and our other Ks-band light curve,
are not sufficiently accurate to enable a similar comparison.

Lastly, it should be noted that even if we are able to accurately estimate the timescale of the periodic repetition
of features in the light curves of SIMP 0136, this may not provide an accurate
estimate of the rotation period of the ultra-cool dwarf.
For Jupiter the rotation period of the planet's magnetosphere, as measured via radio observations,
is different than the rotation period of clouds near the equator \citep{NewburnGulkis73}.
%%% - a factor that has been attributed to the speed of Jupiter's winds.
Although SIMP 0136 has been detected at radio wavelengths \citep{Kao16},
a radio period has yet to be determined for this dwarf.
In conclusion, our photometric period estimate 
of 
\WeightedMeanSIMPPeriod \ $\pm$ \WeightedErrorSIMPPeriodScaled \ hours is
probably best described as 
the rotation period of clouds, or cloud gaps, enveloping 
this L/T transition brown dwarf.

\subsection{The amplitude evolution of SIMP 0136}

\begin{figure}
% \centering
\includegraphics[scale=0.65, angle = 270]{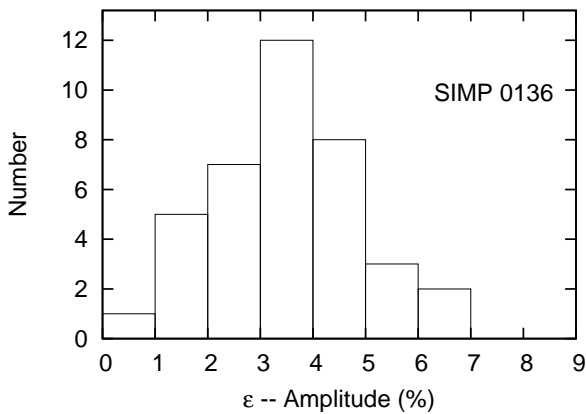}
\caption[]
	{	Histogram of the peak-to-peak amplitudes per complete rotation period of our J-band photometry of SIMP 0136.
		SIMP 0136 displays significant evolution in the peak-to-peak amplitudes 
		of the light curve from rotation period to rotation period.
	}
\label{FigSIMP0136Amplitude}
\end{figure}

\begin{figure}
% \centering
\includegraphics[scale=0.55, angle = 270]{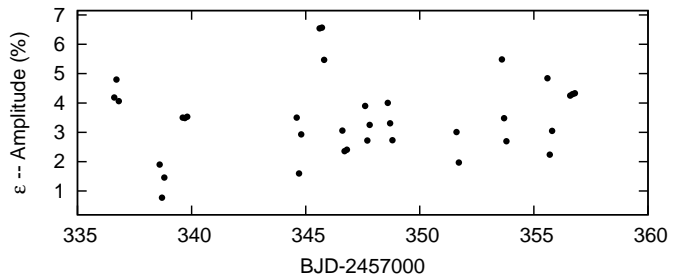}
\caption[]
	{	The peak-to-peak amplitude of variability ($\epsilon$) per complete rotation period for a subset of 
		our J-band photometry
		of SIMP 0136 (from 2015 November, during which we have the most complete nights of data).
	}
\label{FigSIMP0136AmplitudeBJD}
\end{figure}

We also measure the evolution of the SIMP 0136 J-band light curves with time.
Specifically we investigate the 
change in variability amplitude of the SIMP 0136 light curves 
from rotation period to rotation period, and night-to-night.
We use the technique detailed in \citet{CrollTwoMass} that we summarize here.
For each complete rotation period of SIMP 0136 we determine the peak-to-peak amplitude, $\epsilon$,
of the observed variability of that rotation period. To approximate $\epsilon$ we first 
exclude significant outliers (we cut out all data greater than 4 standard deviations from the mean);
then we time-bin the data,
by taking a running average of the photometry every \DataRunningDays \ days (or $\sim$\DataRunningMinutes \ minutes).
We then define $\epsilon$ as the difference between the maximum and minimum points of the time-binned
data observed over each rotation period.
We calculate $\epsilon$ for each complete rotation period
that we have observed; 
%%%%%%%%%%%%%%%%%%
%%%Two different ways of denoting a rotation period.
%a complete rotation period is defined when we have observed a full rotation period between the dashed
%vertical lines of Figure \ref{FigSIMP0136One} \& \ref{FigSIMP0136Two}).
%%%%%%
a complete rotation period is defined when we have observed a full rotation period
after the start of each night of observations (we also exclude $\sim$4 minutes at the start and end of each night 
of observations to avoid possibly, systematically biased photometry).
%%%%%%%%%%%%%%%%%%
We display a histogram of the peak-to-peak amplitude values, $\epsilon$, for each complete rotation period
for our J-band photometry in Figure \ref{FigSIMP0136Amplitude}.
%INSERTHERE CHECK
Our J-band photometry features significant evolution in the light curve from rotation period to rotation period. 
The maximum peak-to-peak amplitude we observe in our J-band photometry is in excess of $\sim$\SIMPMaximumRot\%
(displayed for a few rotation periods in a row on 2015 November 19),
while the minimum is less than $\sim$\SIMPMinimumRot\% (displayed for a handful of rotation periods on 2015 November 12).
The significant evolution we observed in our J-band photometry underscores how important it is to continue performing multiwavelength photometry
of SIMP 0136 and other highly variable L/T transition brown dwarfs to continue to observe the evolution, or lack thereof, in their variability.

Those performing spectrophotometric studies of the variability of L/T transition brown dwarfs should be aware that
significant evolution in the light curves may occur that might drastically change the conclusions one draws on the physical causes of the observed variability,
or the exact ratio of spectral features from the minima to maxima of the observed variability.
%CHECKTHIS
For instance,
the \citet{Apai13} HST/WFC3 spectrophotometric study of SIMP 0136 occurred when SIMP 0136 seemed to display $\sim$5\% J-band variability,
and examined the resulting changes in the near-infrared spectra of SIMP 0136 between the maximum and minimum flux levels observed. 
It seems likely that the changes in HST/WFC3 spectra, or other spectra, of SIMP 0136 may be drastically different when the peak-to-peak amplitudes of variability
are less than \SIMPMinimumRot\% compared to when they are in excess of \SIMPMaximumRot\%.
Therefore, one should be aware that the conclusions reached by \citet{Apai13}, when they observed $\sim$5\% peak-to-peak J-band variability, resulted from an epoch when 
SIMP 0136 displayed larger amplitude variability than on average (Figure \ref{FigSIMP0136Amplitude}; assuming that our photometric monitoring constitutes an accurate representation
of the long-term variability of SIMP 0136).

We also note that the amplitudes of variability from one rotation period to the next are not random for SIMP 0136, but appear to be highly correlated
with the variability amplitude of the previous rotation period. 
In Figure \ref{FigSIMP0136AmplitudeBJD} we present the peak-to-peak amplitudes
per complete rotation period in November 2015 when we collected most of our SIMP 0136 data.
Several distinct clumps are apparent on most of the nights of observations compared to the other nights,
and therefore it appears that for most rotation periods
the amplitude of variability is highly correlated with the variability amplitude from the previous
rotation period. Although some evolution is readily apparent from one rotation period
to the next (in Figures \ref{FigSIMP0136One} \& \ref{FigSIMP0136Two}), 
in general the amplitude of variability is closely correlated with the amplitude from the previous rotation period.
On the other hand we frequently observe significantly different variability amplitudes, or the profile of variability,
from one night's observations to the next; for instance the small amplitude variability, reaching a minimum
of less than
\SIMPMinimumRot\%, on 2015 November 12 is followed by consistent
peak-to-peak variability of $\sim$3\% the next night (2015 November 13).
We note that the variability at the start of the night on 2015 November 12, might signal the possibility of significant evolution
on a timescale of less than a rotation period, as $\sim$2\% variations appear to 
abruptly disappear.
Therefore the timescale for significant evolution of the gaps in clouds that are believed to be driving the variability
on SIMP 0136 appear to be from as short as $<$1 rotation period, and typically not longer than $\sim$10 rotation periods; the variability
of this L/T transition brown dwarf 
seems to frequently evolve on timescales longer than a rotation period, but less than a day.

%INSERTHERE?
%INSERTHERE FURTHER ANALYSIS OF AUTOCORRELATION OF AMPLITUDES WITH TIME.

\subsection{Variability over the Night of SIMP 0136}
\label{SecVariabilityNight}

We note that several of our SIMP 0136 
J-band light curves appear to feature decreasing trends over the night of observations;
the 2015 November 21 and 2015 November 30 observations are prime examples.
We do not know the source of these decreasing trends,
and whether they are astrophysical, systematic or instrumental.
On 2015 November 31 when differential photometry is performed on our reference stars
in the exact same manner as is performed on our target star,
several of the reference stars show decreasing or increasing trends.
However, this is not the case for reference stars on 2015 November 21.
If these trends are not astrophysical, and therefore do not represent a decrease in flux of SIMP 0136 over the course
of a night, then the likely cause of these trends are systematic or instrumental.
Systematic possibilities include that the trends represent imperfect sky subtraction,
or are related to colour effects with airmass of the redder target star, compared to the bluer reference stars (as
we likely observed for Ks-band near-infrared photometry of the M-dwarf GJ 1214; \citealt{CrollGJ}).
We note that such nightly trends are not obviously present in our J-band Perkins/Mimir photometry of the L3.5 dwarf
2MASSW J0036159+182110 \citep{CrollTwoMass}, which is even redder than SIMP 0136 ($J-K$=1.41 for 2MASSW J0036159+182110, compared
to $J-K$=0.89 for SIMP 0136, for which both have typically bluer reference stars).
Also, we frequently observed linear or quadratic trends in the out-of-eclipse light curves for our previous observations
of the near-infrared thermal emission of hot Jupiters, and the transit of a super-Earth
\citep{CrollTrESTwo,CrollTrESThree,CrollWASPTwelve,CrollGJ,CrollRedux},
however these trends were of much smaller magnitude than we observe here.
Therefore, although it is certainly possible that these trends are astrophysical 
(such as may be the case on the night of 2015 November 21),
and represent a decrease in the flux of SIMP 0136 over the course of the night, we cannot rule out the possibility
that these trends are systematic or instrumental in origin (as seems likely for the night of 2015 November 30).

\subsection{Transiting Planets \& Flares}

We further note that our long-term light curves of SIMP 0136 should also allow us to rule out transiting
Earth or super-Earth sized and larger sized planets in the so-called habitable zone of this ultra-cool dwarf.
We do not perform injection and recovery of transiting planet for our SIMP 0136 light curves, as we
do for the ultra-cool dwarf 2MASSW J0036159+182110 \citep{CrollTwoMass}; however, we note that we recover
no obvious signs of a transiting planet during our  
\INSERTNIGHTSSIMP \ nights of photometry of SIMP 0136. In general, super-Earth sized transiting planets should
be readily apparent in our photometry.

Our near-infrared and optical photometry of SIMP 0136 also 
allows us to place a limit on the occurrence rate of flares
exhibited by this L/T transition brown dwarf.
Inspection of the unbinned light curves suggest that we can rule
out flaring events 
lasting thirty minutes or more 
that increase the flux of SIMP 0136 by 10\% or more for 
all $\sim$\TOTALNONOVERLAPPINGHOURS \
non-overlapping hours of photometry that we present here.
More detailed inspection of the light curves could likely present 
considerably more stringent limits on the occurrence of flares.

%FLARES

\section{TVLM 513}
\label{SecAnalysisTVLM}

\begin{figure}
% \centering
\includegraphics[scale=0.45, angle = 270]{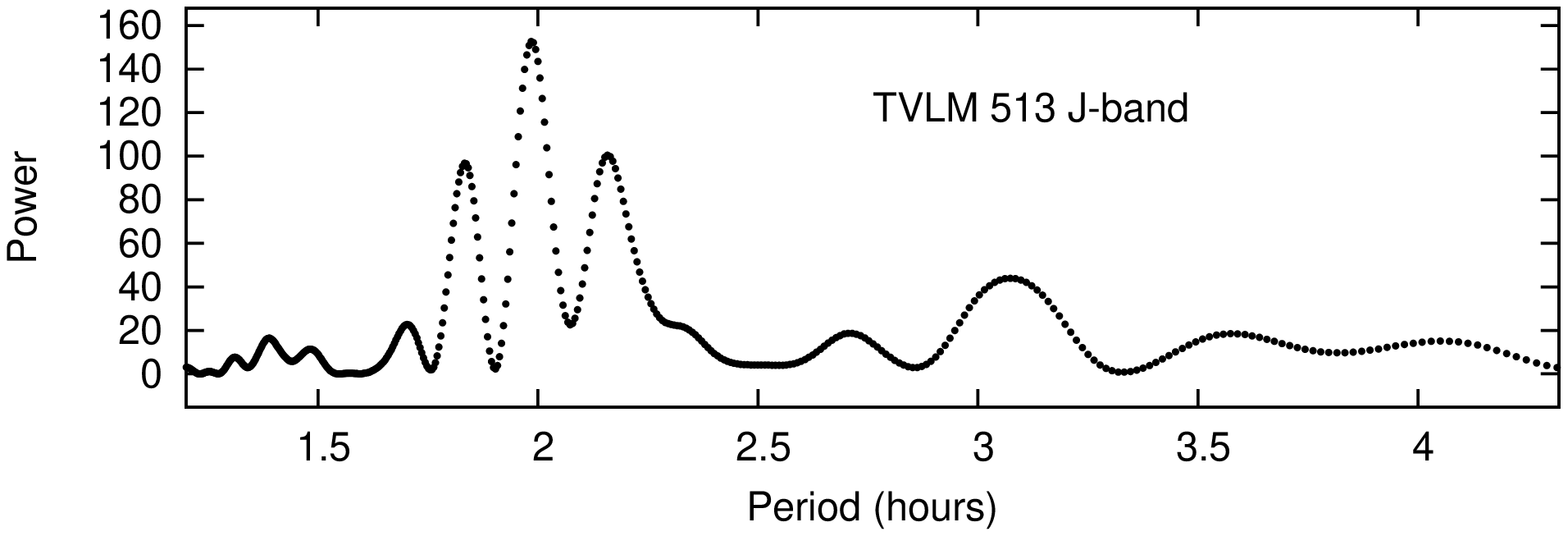}
\includegraphics[scale=0.45, angle = 270]{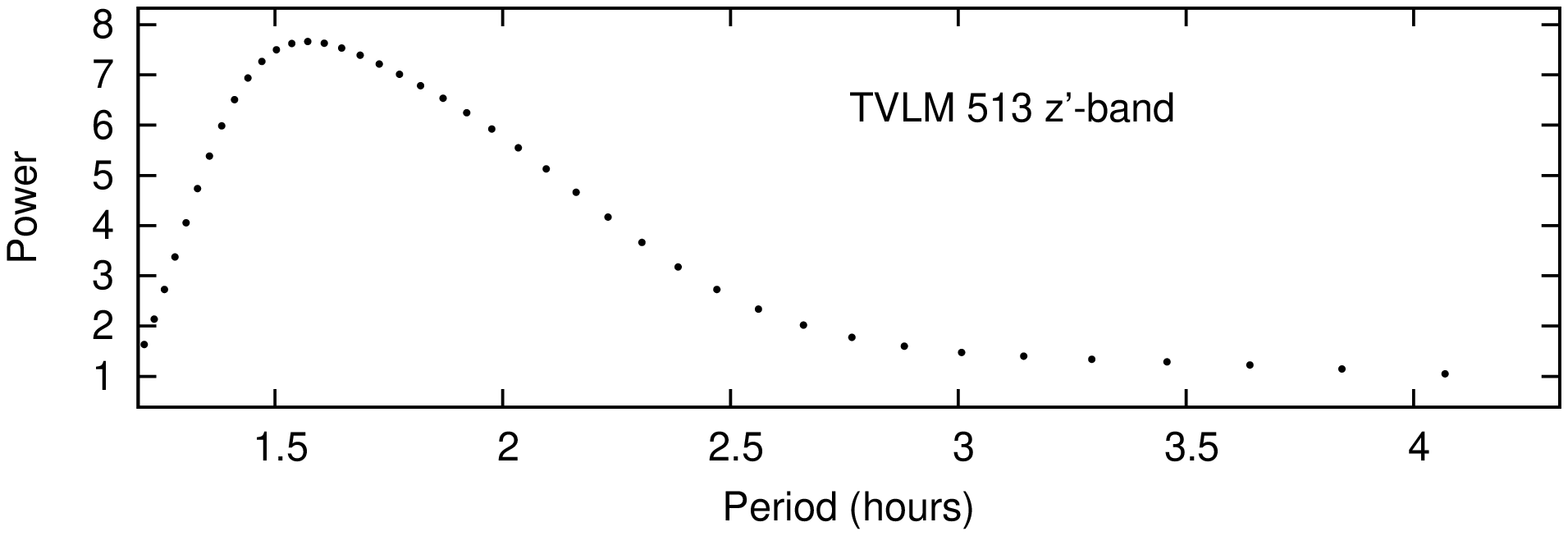}
\includegraphics[scale=0.45, angle = 270]{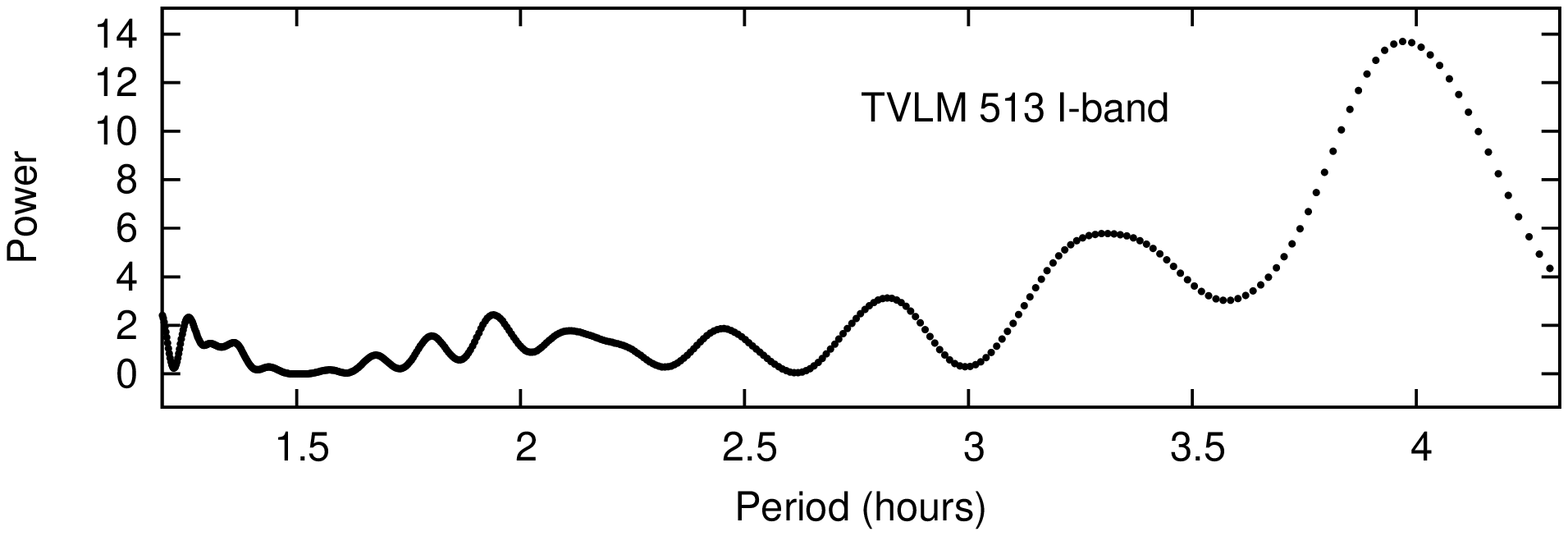}
\caption[]
	{	Lomb-Scargle Periodogram of our photometry of TVLM 513
		in various bands as indicated in the panels 
		(note the different vertical scales on the y-axis).
		Only our J-band photometry displays periodicity
		that is statistically significant; our J-band photometry has a 	
		peak periodogram value of 
		\PeriodHoursTVLMJband \ $\pm$ \PeriodHoursErrorTVLMJband \ hours.
		%and with all our photometry (bottom). We present the peak periodogram values in Table \ref{TablePeriods},
		%which are all approximately $\sim$\PeriodHoursTwoMassZeroZeroThreeSixApprox \ hours.
		
	}
\label{FigTVLM513LombPeriodogram}
\end{figure}

\begin{figure}
% \centering
\includegraphics[scale=0.45, angle = 270]{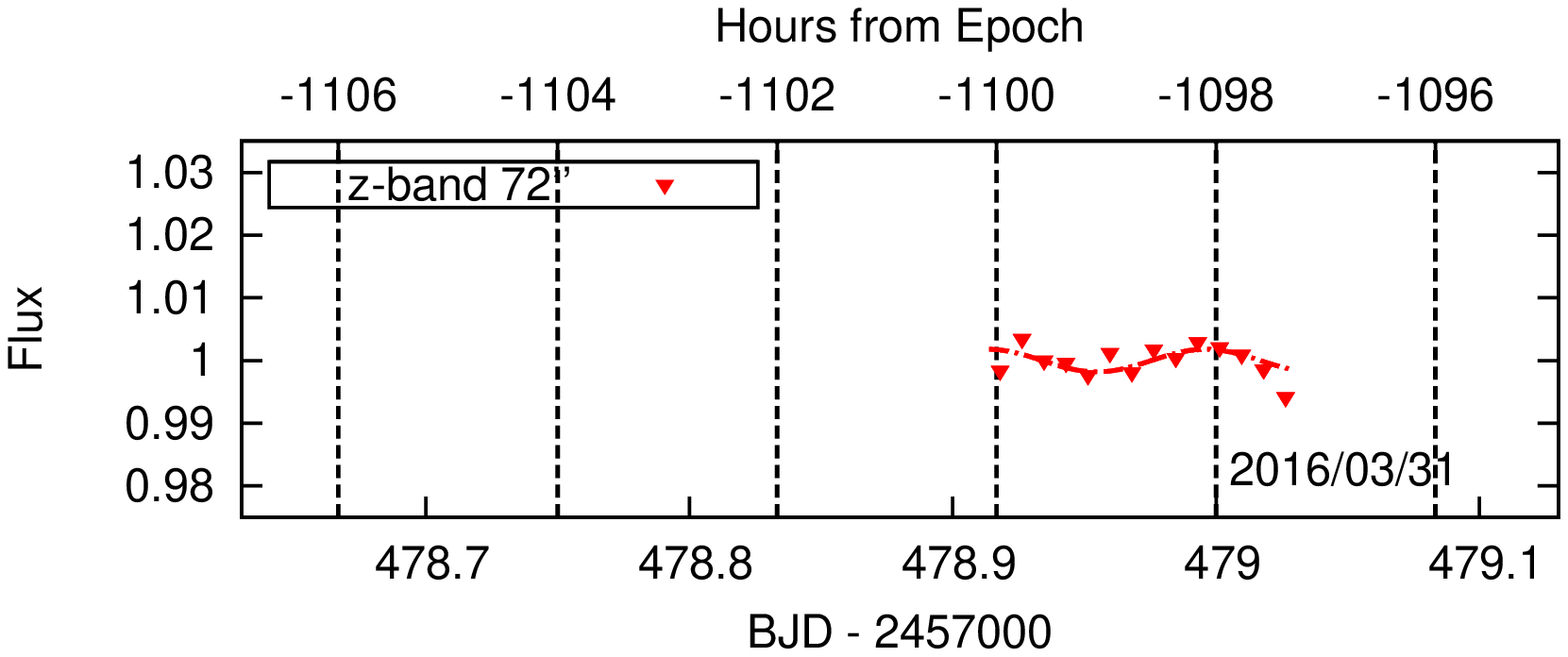}
\includegraphics[scale=0.45, angle = 270]{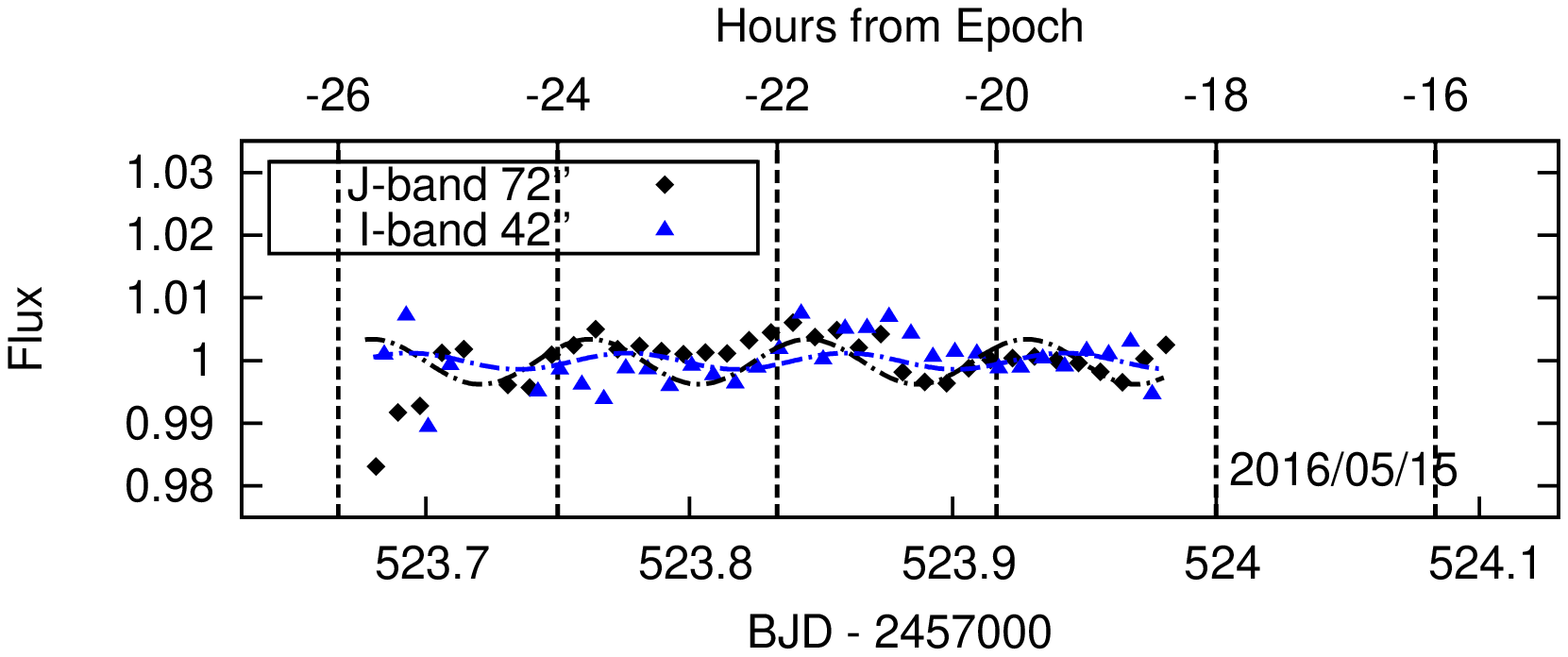}
\includegraphics[scale=0.45, angle = 270]{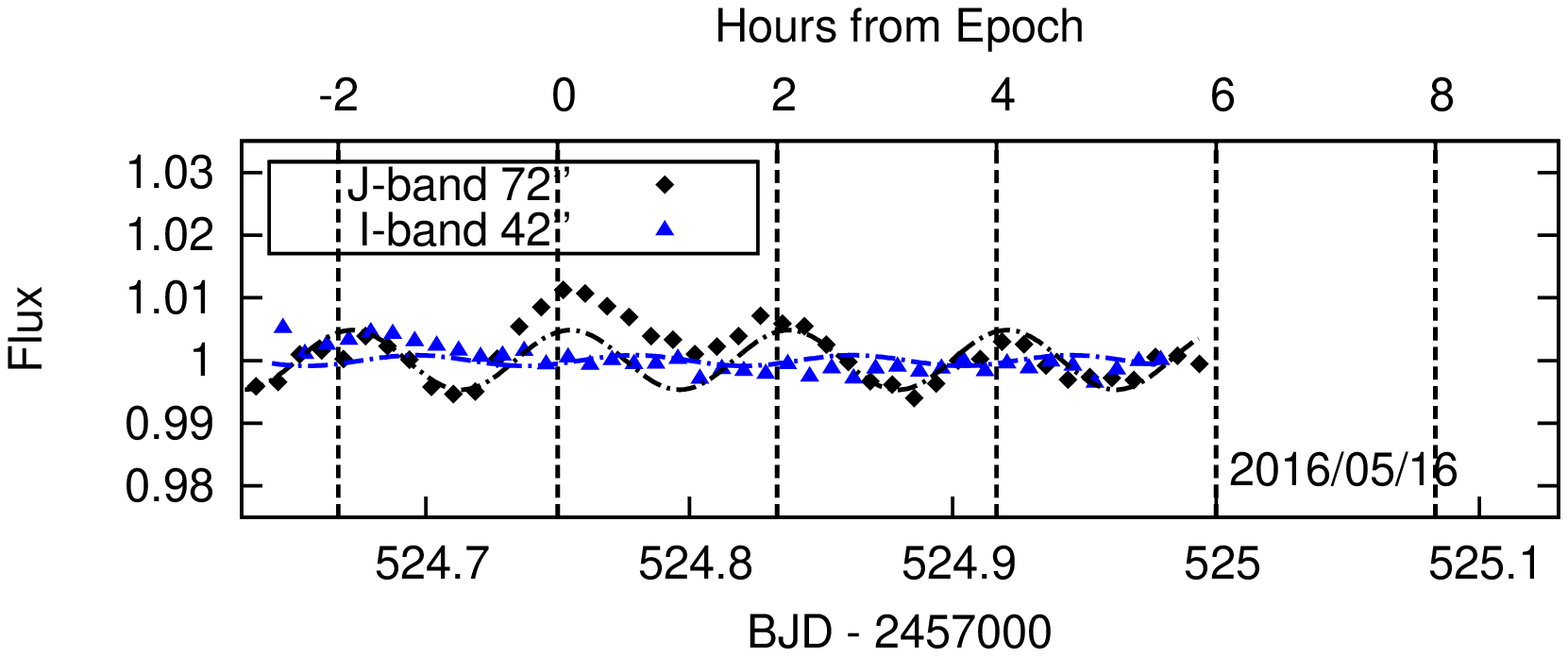}
\caption[]
	{	Sinusoid fits (dotted lines) to the simultaneous, multiwavelength
		Perkins, \& Hall DCT photometry of TVLM 513 in the J and I-bands.
		The colour of the dot-dashed lines corresponds to the wavelength of observations
		as denoted by the colours of the data-points 
		as indicated in the legend.
		The figure format is otherwise identical to Figures \ref{FigTVLM}.
		For plotting clarity we remove linear trends from our J-band data-sets. % that we do not believe to be astrophysical in nature
	}
\label{FigTVLMJointSinusoids}
\end{figure}

\begin{deluxetable}{ccccc}
\tablecaption{Sinusoid Fits to Individual Photometric Light Curves of TVLM 513}
\tablehead{
\colhead{Date}	& \colhead{Telescope}	& \colhead{Band} 	& \colhead{Peak-to-Peak}		&  \colhead{Phase}		\\
\colhead{(UTC)}	& \colhead{}		& \colhead{} 		& \colhead{Amplitude $A$ (\%)}		&  \colhead{($\phi$)}	\\
}
\centering
\startdata
2016/03/31	& Perkins		& z'		& \SinusoidPeaktoPeakAmplitudePercentageTVLMSixteenMarchThreeOnezband \ $\pm$ \SinusoidPeaktoPeakAmplitudePercentageErrorTVLMSixteenMarchThreeOnezband		& 	\SinusoidPhaseTVLMSixteenMarchThreeOnezband \ $\pm$ \SinusoidPhaseErrorTVLMSixteenMarchThreeOnezband	 \\	%
\hline
2016/05/15	& Perkins		& J		& \SinusoidPeaktoPeakAmplitudePercentageTVLMSixteenMayFifteenJband \ $\pm$ \SinusoidPeaktoPeakAmplitudePercentageErrorTVLMSixteenMayFifteenJband		& 	\SinusoidPhaseTVLMSixteenMayFifteenJband \ $\pm$ \SinusoidPhaseErrorTVLMSixteenMayFifteenJband	 \\	%
2016/05/15	& Hall			& I		& \SinusoidPeaktoPeakAmplitudePercentageTVLMSixteenMayFifteenIband \ $\pm$ \SinusoidPeaktoPeakAmplitudePercentageErrorTVLMSixteenMayFifteenIband		& 	\SinusoidPhaseTVLMSixteenMayFifteenIband \ $\pm$ \SinusoidPhaseErrorTVLMSixteenMayFifteenIband	 \\	%
\hline
2016/05/16	& Perkins		& J		& \SinusoidPeaktoPeakAmplitudePercentageTVLMSixteenMaySixteenJband \ $\pm$ \SinusoidPeaktoPeakAmplitudePercentageErrorTVLMSixteenMaySixteenJband		& 	\SinusoidPhaseTVLMSixteenMaySixteenJband \ $\pm$ \SinusoidPhaseErrorTVLMSixteenMaySixteenJband	 \\	%
2016/05/16	& Hall			& I		& \SinusoidPeaktoPeakAmplitudePercentageTVLMSixteenMaySixteenIband \ $\pm$ \SinusoidPeaktoPeakAmplitudePercentageErrorTVLMSixteenMaySixteenIband		& 	\SinusoidPhaseTVLMSixteenMaySixteenIband \ $\pm$ \SinusoidPhaseErrorTVLMSixteenMaySixteenIband	 \\	%

\enddata
% \tablenotetext{c}{The duration indicates the time between the first and last observation of the evening, and does not take into account gaps in the data due to clouds, humidity or poor weather.}
\label{TableSinusoidIndividualLightCurvesTVLM}
\end{deluxetable}

Our simultaneous J and I-band TVLM 513 light curves display prominent J-band variability.
We present Lomb-Scargle periodograms of our TVLM 513 light curves in 
Figure \ref{FigTVLM513LombPeriodogram}.
For our J-band photometry the 
peak periodogram value is \PeriodHoursTVLMJband \ $\pm$ \PeriodHoursErrorTVLMJband \ hours.
For our I and z'-band photometry the peak periodicity is small enough
that we do not believe it
to be statistically significant (as we indicate below).
The J-band peak periodogram value of \PeriodHoursTVLMJband \ $\pm$ \PeriodHoursErrorTVLMJband \ hours is 
very similar to the previously claimed rotation period of this dwarf:
\citet{Hallinan07}, \citet{Lane07}, \citet{Littlefair08} \& \citet{Harding13} detected
$\sim$2 hour radio and optical variability from TVLM 513 and inferred that 
this was the rotation period of this M-dwarf.

For our TVLM 513 light curves to determine the amplitude
of the J-band variability, and to place limits
on the I and z'-band variability,
we fit the individual light curves
with sinusoidal fits utilizing a period of \PeriodHoursTVLMJband \ hours.
For each individual band we fit for the peak-to-peak amplitude, $A$, phase, $\phi$,
and a vertical offset, $y$. 
$\phi$ is defined from 0 - 1 and
$\phi$ = 0 denotes the flux maximum of the sinusoid.
The results are given in Table \ref{TableSinusoidIndividualLightCurvesTVLM}, and presented in Figure \ref{FigTVLMJointSinusoids}.
To enhance plotting clarity we subtract out linear trends from our J-band data-sets in Figure \ref{FigTVLMJointSinusoids}.
%For 2015/05/16 most of the reference stars do not show trends, but some do.
%2015/05/15 bad clouds at start of the night, and then we occasionally so trends.
Similarly as for SIMP 0136 (see Section \ref{SecObs}), it is unclear whether these trends are astrophysical or systematic/instrumental in nature.

Our J-band photometry on both nights of multiwavelength photometry displays
statistically significant variability with a $\sim$\PeriodHoursTVLMJband \ hour period,
while the simultaneous 
I-band photometry does not display statistically significant variability.
Similarly, our z-band photometry on 2016 March 31 does not display statistically significant variability with a 
$\sim$\PeriodHoursTVLMJband \ hour period.
We also fit our I and z'-band light curves with the peak periodogram values from our Lomb-Scargle analysis in 
Figure \ref{FigTVLM513LombPeriodogram} -- we confirm that the variability in our I and z'-band light curves
at these periods are not significant at the 3$\sigma$ level.

The fact that we detect prominent $\sim$0.7 - 0.9\% J-band variability while the I-band variability is significantly
less during our simultaneous, multiwavelength photometry of TVLM 513 suggests that clouds or aurorae are responsible for the variability
displayed by this M9 dwarf. Stronger near-infrared variability than optical variability is inconsistent with starspot driven variability
for starspots that are either hotter or cooler than the TVLM 513 photosphere.

%CHECKHERE could comment about there being no huge absorption line in I-band that makes the flux of the two consistent with one another
%according to the spectrum...

We also note that our lack of detection of I-band variability, with a 3$\sigma$ upper 
limit on sinusoidal variability on our two nights
of multiwavelength photometry of $\sim$\SinusoidPeaktoPeakAmplitudePercentageThreeSigmaTVLMSixteenMayFifteenIband\%
on 2016/05/15 and $\sim$\SinusoidPeaktoPeakAmplitudePercentageThreeSigmaTVLMSixteenMaySixteenIband\% 2016/05/15,
is inconsistent with previous detections of obvious variability for TVLM 513.
Our limits on I-band variability are less than the $\sim$10 mmag variability detection of \citet{Lane07} from
12 hours of monitoring, and inconsistent with most of the nights of
I and i'-band monitoring from \citet{Harding13} with peak-to-peak amplitudes of $\sim$0.6 - 1.2\%.
Although the wavelength regions do not completely overlap, we also note that our variability amplitudes are considerably
less than the 3-4\% variability detection of \citet{Littlefair08} from a few hours of
g' and i'-band monitoring. 
Therefore, it appears clear that the amplitude of optical variability from TVLM 513
varies considerably from epoch to epoch.

\section{Discussion \& Conclusions}

We have returned long-term, ground-based photometry of two ultra-cool dwarfs:
the T2.5 L/T transition brown dwarf SIMP 0136 and the M9 radio-active, ultra-cool dwarf TVLM 513.

Our 
\INSERTNIGHTSSIMP \ nights of ground-based,
near-infrared photometry of SIMP 0136,
including
\INSERTNIGHTSSIMPJBAND \ nights of J-band photometric monitoring spread out over \INSERTSPREADDAYSSIMPJBAND\ days,
allow us to observe how the variability of SIMP 0136
evolves from rotation period to rotation period, night-to-night, \& week to week.
We estimate the rotation period of SIMP 0136 as
\WeightedMeanSIMPPeriod \ $\pm$ \WeightedErrorSIMPPeriodScaled \ hours.
We do not detect any significant deviations in the photometric period from night-to-night, and therefore constrain
the frequency and magnitude of obvious differential rotation;
if differential rotation is present on SIMP 0136 and drives the variability 
with a different period for an entire night during our observations,
then 
the period must deviate by less than $\sim$\DiffRotationPercentage\% from our estimate of the rotation period of SIMP 0136.

The 
peak-to-peak amplitude 
displayed by SIMP 0136
in our J-band light curves evolves from greater than \SIMPMaximumRot\% to less than \SIMPMinimumRot\% in the space of 
just a few days.
Longer-term spectrophotomeric studies, and/or Doppler imaging studies,
of individual L/T transition brown dwarfs are crucial in
determining whether the rapid evolution of the variability that we observe 
from rotation period to rotation period
is due to
the size of thinner-cloud (or cloud-free) regions growing or shrinking, or
that the characteristics of the clouds (e.g. the thickness, height, or type of clouds)
in thinner-cloud regions of approximately constant size are changing.
Although spectrophotomeric studies to date have suggested that that the variability observed in L/T transition brown dwarfs over approximately
a single rotation period
are well modeled by only two surfaces consisting of thick clouds, and warmer, thinner clouds \citep{Apai13,Buenzli15},
our observations indicate that this comparison may only be representative of the spectral 
characteristics at a single ephemeral snapshot in time, and may not represent the full continuum of spectral characteristics.

% The rapid evolution that we observe suggests that those performing spectrophotometric observations
% of variable brown dwarfs should be aware that the comparison between the spectra obtained at maximum flux and minimum flux
% of a single rotation period,
% may represent the spectral 
% characteristics of a single, 

The timescale for significant evolution of the SIMP 0136 light curve appears to be as short as a rotation period,
and as long as approximately a day.
This is arguably the first time that the timescale for the evolution of the light curve
of a L/T transition brown dwarf has been robustly measured. 
With the possible exception of Luhman-16, for other highly variable L/T transition brown dwarfs
there have yet to be published a sufficient number of multi-night light curves
in a single observing band for the timescale of evolution to be robustly measured 
(e.g. \citealt{Radigan12} for 2MASS J21392676+0220226).
As for Luhman-16, 
\citet{Gillon13} published observations suggesting strong night-to-night variations,
but it is less clear if the binary exhibits significant rotation period to rotation period variations. 
Nonetheless, determining a precise timescale for 
the light curve evolution of a single member of the Luhman-16
binary brown dwarf system is challenging
due to the fact that both L/T transition binary members are variable \citep{Biller13,Buenzli15b}.
However, several studies have indicated that the majority of the variability from the Luhman 16 system 
originates from Luhman 16B \citep{Gillon13,Burgasser14,Buenzli15};
if this is the case, then the significant correlation
that we observe of the peak-to-peak amplitudes from one rotation to the next 
for SIMP 0136
suggests that the 
timescale for significant evolution of the light curve might be longer for SIMP 0136 than for Luhman-16B.

For the radio-active, ultra-cool dwarf TVLM 513, our \INSERTNIGHTSSIMULTANEOUSTVLM \ nights of simultaneous, ground-based photometry 
display 
obvious J-band variability, without accompanying obvious I-band variability of similar amplitude.
This confirms that the variability of TVLM 513 
likely arises from clouds or aurorae, rather than starspots.

We encourage further monitoring of these two intriguing ultra-cool dwarfs, and other ultra-cool dwarfs,
to better constrain the long-term
evolution, or lack thereof, of their variability.

\acknowledgements

We thank Peter Williams for a careful reading of this manuscript and comments that improved
this manuscript.
We also thank Saul Rappaport \& Adam Burgasser for helpful discussions that
improved this manuscript.

Some of this effort was supported by the NASA Exoplanet Research Program (XRP) under Grant No. NNX15AG08G issued through the
Science Mission Directorate, and by the National Science Foundation under Grant No. 1359205.

\end{document}